\documentclass{aa}

\usepackage{graphicx}
\usepackage{txfonts}
\usepackage{xcolor}

\begin{document} 
    \titlerunning{Merger-induced disturbance in clusters}
 \authorrunning{Kong \& Dell'Antonio}

   \title{Merger-induced disturbance and temporal signatures in galaxy clusters: a combined phase space and photometric analysis}

   \author{Chuiyang Kong
          \inst{1}\fnmsep\thanks{Corresponding author: chuiyang\_kong@brown.edu}
          \and
          Ian Dell’Antonio\inst{1}}

   \institute{Department of Physics, Brown University, 182 Hope Street, Providence, RI 02912, USA\\
              \email{chuiyang\_kong@brown.edu, ian\_dell'antonio@brown.edu}}

   \date{Received September 24, 2025}
 
  \abstract 
   {We present a physically interpretable framework to quantify dynamical disturbances in galaxy clusters using projected two-dimensional phase-space information. Based on the TNG-Cluster simulation, we construct a disturbance score that captures merger-driven asymmetries through features such as velocity dispersion and Gaussian Mixture Model (GMM) peak fitting, which captures asymmetries indicative of dynamical disturbance. All features are derived from observable quantities and are intended to be measurable in future surveys.

To enable observational application, we adopt a simplified estimator using aperture mass map statistics as a mass ratio proxy in TNG300-1, and validate its performance with weak lensing data from The Local Volume Complete Cluster Survey (LoVoCCS). While phase-space diagnostics reveal merger-driven asymmetries, they are not sensitive to whether the secondary progenitor is infalling or receding, and thus cannot distinguish future mergers from past mergers. To address this, we incorporate the star formation rate (SFR) from TNG-Cluster and propose the blue galaxy fraction as a promising observational tracer of merger timing.

Finally, we construct mock Chandra X-ray images of TNG-Cluster halos at redshift $z=0.2$, and find that the offset between the X-ray peak and the position of the most massive black hole (used as a proxy for the Brightest Cluster Galaxy, BCG) correlates with our disturbance score, serving as a consistency check. We also perform case studies using LoVoCCS observational data, correlating the blue galaxy fraction with disturbance scores derived from the eROSITA morphology catalog.}


   \maketitle
%

\section{Introduction}

Galaxy clusters, the most massive gravitationally bound structures in the universe, serve as critical tracers of structure formation and laboratories for a wide range of astrophysical processes. Cluster mergers, especially major mergers with mass ratios above 0.1 (i.e., where the secondary progenitor contains at least 10\% of the mass of the primary), are among the most energetic events in the current universe. These mergers release gravitational energies exceeding $10^{64}$ erg, driving shocks and turbulence that heat the intracluster medium (ICM), a diffuse, X-ray–emitting plasma that fills the space between galaxies \citep{clustermergers, 2019_intro_2}.

Beyond disturbing the ICM, mergers may also influence the star formation activity of cluster galaxies. However, whether they trigger starbursts or accelerate quenching remains uncertain. Both observations and simulations have reported a wide range of outcomes, including enhanced star formation (e.g., \cite{sim_burst_Contreras_Santos_2022, obs_burst_2023_Aldas}) and merger-induced quenching (e.g., \cite{obs_quench_Deshev_2017, obs_quench_roberts2024mergershocksenhancequenching}). Other studies find no significant global effect beyond localized starbursts (e.g., \cite{obs_no_significant_Rawle_2014}), or suggest a two-phase sequence of initial enhancement followed by suppression (e.g., \cite{obs_combined_Stroe_2015, obs_combined_Sobral_2015}). These mixed results underscore the complexity of merger-driven galaxy evolution.

Despite their astrophysical significance, cluster mergers remain difficult to identify and characterize observationally. A variety of diagnostic tools have been developed, each targeting distinct physical signatures. The common goal is to detect asymmetries and structural disturbances indicative of merger activity. In X-ray imaging, morphological indicators, such as symmetry, peakiness, alignment, concentration, and multipole moments, have been widely used to distinguish dynamically relaxed clusters from disturbed systems (e.g., \cite{Mantz_2015, erosita_morphology_cat_Sanders_2025}). The projected offset between the X-ray peak and the position of the brightest cluster galaxy (BCG) is another commonly used proxy, based on the expectation that mergers displace the hot gas relative to the central galaxy (e.g., \cite{BCGoffsets_1984_Jones, BCGoffsets_Katayama_2003, BCGoffsets_Sanderson_2009}). Similarly, the offset between the peaks of gravitational lensing mass maps and X-ray emission has also been proposed as an indicator of dynamical disturbance \citep{BCGoffsets_Poole_2006}. The Bullet Cluster (1E 0657-56, \cite{2006_bullet_cluster}) and its analogues (e.g. \cite{bullet_cluster_2008_analogues2, bullet_cluster_2024_analogues1}) provide compelling examples of such separation due to mergers.

More detailed X-ray morphology, such as the presence of cold fronts (e.g., \cite{cold_fronts_Ghizzardi_2010}) and shock edges (e.g., \cite{shock_Botteon_2018}), can also serve as evidence of merger activity when available. In the radio regime, diffuse synchrotron emission such as radio relics provides complementary signatures of merger-driven shocks and turbulence (e.g., \cite{combined_radio_Larrondo_2018, Radio_Golovich_2019, Radio_Wilber_2019}). These multi-wavelength indicators are often combined to improve the robustness of merger identification.

In the optical regime, mergers may appear as multiple peaks in galaxy number density or kinematic substructures in velocity space. Photometric observations can capture spatial asymmetries (e.g., \cite{wen2024cataloguemergingclustersgalaxies}), while spectroscopic measurements, such as those using the H$\alpha$ line, provide dynamical and star formation information. However, obtaining full-sky, multi-wavelength cluster observations remains challenging, making the robust identification of mergers difficult.

While each method offers valuable insight, the identification of merging systems is highly definition-dependent in large observational samples and provides limited information about the timing of merger events. In contrast, cosmological simulations offer full knowledge of each cluster’s merger history, enabling detailed tracking of dynamical evolution and structural disturbance. Many efforts have leveraged simulations to study merger-driven phenomena (e.g. \cite{ZuHoneGalaxyClusterMergerCatalog}, \cite{Arendt_2024}, \cite{lee2024radiorelicsmassivegalaxy}), though the use of phase-space distributions, defined by projected positions and line-of-sight velocities, remains relatively uncommon (e.g. \cite{phase_space_van_der_Jagt_2025}). When combined with star formation histories, color evolution, and synthetic observables such as X-ray morphology or mass maps, such simulation-based approaches offer a powerful testbed for developing generalizable frameworks to trace merger-driven disturbance.

In this work, we aim to develop a physically interpretable framework for characterizing merger-driven disturbance in galaxy clusters. Using the TNG-Cluster simulation \citep{nelson2024introducingtngclustersimulationoverview}, we extract features from the phase-space distribution of member galaxies and train a machine learning model to predict a continuous disturbance score. The model is supervised using the true merger history of each cluster, allowing us to construct a score that captures the structural impact of major mergers in a time-sensitive manner.

Since the phase-space machine learning model lacks the ability to distinguish between future and past mergers, corresponding to infalling versus outgoing second progenitors, we further analyze the evolution of SFR across merger events and examine the blue galaxy fraction as a potential time-sensitive tracer. We find that the disturbance score correlates with both the curvature of star formation histories and the blue fraction in a way that reflects merger timing. As a consistency check, we compare the score with BCG offsets measured from mock X-ray maps.

To connect our framework to observable quantities, we develop a method to estimate mass ratios from aperture mass maps. This estimator is validated using both the TNG300-1 simulation and LoVoCCS observational data \citep{LoVoCCSI, fu2024lovoccsiiweaklensing, AAS_Englert_2025, AAS_Fu_2025}. We further apply our framework to several LoVoCCS clusters as case studies.

This paper is organized as follows. In Section 2, we describe the simulations and methodology. Section 3 presents the machine learning model, the evolution of star formation rate (SFR), the behavior of blue galaxy fractions, the correlation between BCG offsets and our disturbance score, and our aperture mass ratio estimator. In Section 4, we discuss the grouping strategy for model training, the impact of projection choices, efforts to mitigate richness dependence, and case studies applying our method to LoVoCCS clusters. We summarize our conclusions in Section 5.

\section{Data and methods}
\subsection{The TNG-Cluster simulation}
TNG-Cluster \citep{nelson2024introducingtngclustersimulationoverview} is a spin-off project of the IllustrisTNG suite \citep{nelson2021illustristngsimulationspublicdata}, with particular emphasis on high-mass clusters beyond those captured in TNG300. To improve statistics for massive systems, the parent box of TNG-Cluster is significantly larger than that of TNG300, reaching up to 1 Gpc, and includes 352 high-mass clusters ($> 10^{14}~M_{\odot}$) via zoom-in simulations. The zoom-in regions feature resolutions comparable to TNG300-1, with a dark matter particle mass of $6.1 \times 10^{7}~M_{\odot}$ and a mean baryonic cell mass of $1.2 \times 10^{7}~M_{\odot}$.

We use the 352 primary zoom-in halo targets in TNG-Cluster, along with their main progenitors, as our main sample. We trace them from snapshots 72 to 99, ranging from redshift 0.4 to 0, extract features from their phase-space information, and track the evolution of their SFR and blue galaxy fraction. We also generate mock X-ray images at redshift 0.2. Further details are provided in the following sections.

\subsection{The TNG300-1 simulation}

In addition to the limited number of samples with large mass ratios, we also utilize data from TNG300-1. TNG300-1 is part of the IllustrisTNG suite and has been introduced in a series of papers \citep{nelson2021illustristngsimulationspublicdata, Pillepich_2017, Nelson_2017, Naiman_2018, Marinacci_2018}. In TNG300-1, the dark matter particle mass is $0.9 \times 10^{7} M_{\odot}$, and the mean baryonic cell mass is $1.1 \times 10^{7} M_{\odot}$.

We focus on halos more massive than $10^{13} M_{\odot}$ at redshift 0.08 to validate our rapid method for estimating the mass ratio. In both TNG-Cluster and TNG300-1 analyses, we adopt the same cosmological parameters used throughout the TNG simulations, consistent with the Planck 2015 results \citep{Planck2016}: $\Omega_{\Lambda,0} = 0.6925$, $\Omega_{m,0} = 0.3075$, $\Omega_{b,0} = 0.0486$, $\sigma_8 = 0.8159$, $n_s = 0.9667$, and $h = 0.6774$. These values are defined within the astropy.cosmology module \citep{The_Astropy_Collaboration_2022, Planck2016}.

\subsection{LoVoCCS data}

The Local Volume Complete Cluster Survey (LoVoCCS) is an ongoing program aiming to observe nearly one hundred low-redshift, X-ray–luminous galaxy clusters using the Dark Energy Camera (DECam). Aperture mass maps and weak-lensing–based mass estimates for a subset of these clusters have already been produced through the LoVoCCS pipelines, as described in a series of papers \citep{LoVoCCSI, fu2024lovoccsiiweaklensing, AAS_Englert_2025, 2025ApJL_Englert_2025}.

We perform the proxy analysis with clusters covered by \cite{fu2024lovoccsiiweaklensing}. This serves as an initial attempt to connect our simulation-based framework to observational data across six bands (u, n, g, r, i, z), where we make use of u- and g-band photometry.

\subsection{Phase-space feature extraction}
\label{sec: Phase-space Feature Extraction}

We extract all subhalos associated with the 352 zoom-in targets using the group catalogs of the TNG-Cluster simulation. To define the projected radius on the 2D plane perpendicular to the line of sight, we first compute the projected positions $(x_n,y_n)$ of all subhalos and subtract their mean position $\boldsymbol{\mu} = (\bar{x},\bar{y})$. From these centered coordinates we construct the $2\times2$ covariance matrix.
\begin{equation}
C = \frac{1}{N}\sum_{n=1}^N
  (\boldsymbol{x}_n-\boldsymbol{\mu})(\boldsymbol{x}_n-\boldsymbol{\mu})^{\mathsf T},
\end{equation}
where $\boldsymbol{x}_n$ is the position vector of each subhalo, and $\boldsymbol{\mu}$ is their mean position. We then perform a Principal Component Analysis (PCA) of $C$ using scikit-learn \citep{pedregosa2018scikitlearnmachinelearningpython}. The eigenvectors of $C$ give two orthogonal directions of maximum and minimum variance of the subhalo distribution, and the corresponding eigenvalues give the variance (squared dispersion) along these directions. The eigenvector associated with the largest eigenvalue defines the projected major axis of the halo; we rotate the coordinate system so that this axis lies along the $x$–direction and define the projected radius of each subhalo as its coordinate along this axis. For velocity, we retain only the component along the line of sight, perpendicular to the projected plane, consistent with what is accessible in real observations. We further use the same covariance matrix $C$ to characterize the projected halo shape. Its eigenvalues $\lambda_1 \ge \lambda_2$ represent the variance along the major and minor principal axes, and their ratio $\lambda_2/\lambda_1$ provides a measure of the minor-to-major axis ratio.

To further characterize the phase-space structure, we fit Gaussian Mixture Models (GMMs) using scikit-learn \citep{pedregosa2018scikitlearnmachinelearningpython}
to the projected radius--velocity distribution of all subhalos associated with each halo. From the best-fitting models, we extract features such as the mean projected radius, mean line-of-sight velocity, and the subhalo count of each component. We also record the maximized likelihoods of the $K$-component fits ($K=1,2$) and use them to compute the Bayesian Information Criterion (BIC; \citealt{1978_BIC_def}),
\begin{equation}
\label{eq:bic_eq}
    \mathrm{BIC}_K=-2\ln\hat{\mathcal{L}}_K+k_K\ln N,
\end{equation}
where $\hat{\mathcal{L}}_K$ is the maximum likelihood of the best-fit $K$-component model, $N$ is the number of members used in the fit, and $k_K$ is the number of free parameters.

In addition, we include the mass ratio between the central subhalo and the most massive satellite galaxy within each friends-of-friends (FOF) halo, as identified by the group catalogs.

Our final dataset includes the 352 primary zoom-in halos and their main progenitors from redshift $z = 0.4$ to $z = 0$, corresponding to snapshots 72 to 99. For each halo, we extract phase-space information along all three Cartesian projections. To increase the number of training samples, we treat each projection as an independent input. To avoid data leakage, we test different grouping strategies, which are further discussed in the results and discussion sections.

\subsection{Merger score definition}
\label{sec: Merger Score Definition}
We define the merger score as a time-weighted cumulative contribution from all major merger events associated with a given halo, using an exponentially decaying function of their time separation from the current snapshot:

\begin{equation}
\label{eq:merger_score}
\text{score}_{\text{merger}} = \sum_{\text{mergers}} \exp\left(-\frac{\Delta t}{\tau}\right),
\end{equation}

where $\Delta t$ is the time interval between the current snapshot and the merger event, and $\tau$ is a characteristic decay timescale. Recent mergers (small $\Delta t$) contribute more significantly, while earlier events are exponentially suppressed.

To probe dynamical states across time, we define three versions of the score: the past-merger score, computed by summing only mergers that occurred before the current snapshot; the future-merger score, based on mergers that will occur after it; and the full-merger score, which includes all merger events and equals the sum of the past and future scores when evaluated using the same decay timescale $\tau$.

We consider only major mergers, defined as interactions between a primary halo (i.e., one of the 352 TNG-Cluster zoom-in targets with $M_{500c} > 10^{14}~M_{\odot}$ at redshift 0) and a secondary halo with $M_{\text{halo}} > 10^{13}~M_{\odot}$. Halo masses are taken from the Group\_M\_Crit500 field in the TNG-Cluster group catalogs, corresponding to the mass enclosed within a sphere whose mean density is 500 times the critical density of the universe at the redshift of the halo. Merger information is obtained from the catalog of \cite{lee2024radiorelicsmassivegalaxy}.

\subsection{Star formation and blue galaxy fraction statistics}
\label{sec: Star Formation and Blue Galaxy Fraction Statistics}
We trace the 352 primary zoom-in halos in TNG-Cluster along their main progenitor branches back to redshift 0.4, using the group catalogs to identify their member galaxies. For each FOF halo, we calculate the average star formation rate (SFR) of its member subhalos. To capture merger-induced SFR variations, we analyze SFR evolution within a fixed symmetric 2 Gyr time window centered on each snapshot (1 Gyr before and 1 Gyr after; see Section \ref{sec:Photometric Differentiation of Merger Stages} for details). Specifically, we normalize the SFR of a given halo at a particular snapshot by the average SFR of that halo over this surrounding time window.

To quantify the time evolution, we compute the curvature of the normalized SFR history using the following estimator:
\begin{equation}
\label{sfr_curvature}
\Delta_{\mathrm{SFR}} = 2 \times \mathrm{SFR}_{\mathrm{snap}} - \mathrm{SFR}_{\mathrm{start}} - \mathrm{SFR}_{\mathrm{end}}.
\end{equation}
To estimate the blue galaxy fraction in each FoF halo, we extract absolute magnitudes from the Subhalo\-Stellar\-Photometrics field in the group catalogs. Apparent magnitudes are computed based on redshift. To avoid the regime where observational band errors become substantial, we impose a magnitude cut at $u, g < 23, 25$, and apply the same cut to the simulations for consistency. We define galaxies as “blue” if their $u - g$ color is less than or equal to 0.3. We adopt this relatively strict color cut to avoid overestimating the blue fraction at redshifts near 0, where fainter member galaxies can enter the sample due to the apparent magnitude cut.

\subsection{X-ray image generation}
\label{sec:xray_image_generation}
We select the progenitors of the 352 primary zoom-in halos from TNG-Cluster at redshift $z = 0.2$ (snapshot 84) and download their cutout files, which contain only the high-resolution particles associated with these halos. We adopt $z = 0.2$ as our reference redshift for practical reasons: the hot gas can be identified using the GFM\_CoolingRate field, which is only stored at a small set of snapshots in TNG-Cluster (e.g.\ $z = 0.0$, 0.1, 0.2, 0.3), and the merger trees at $z = 0.2$ still provide a long enough future time span ($\gtrsim 1\,\mathrm{Gyr}$ after this snapshot) to track mergers for our choice of $\tau = 1\,\mathrm{Gyr}$ in the merger scores.

X-ray photon maps are generated using pyXSIM \citep{pyXSIM} and SOXS \citep{SOXS}, with the instrument configuration set to chandra\_acisi\_cy22, assuming an exposure time of 2000 ks and an
effective collecting area of 600 cm$^2$. The Chandra ACIS-I array has a field of view of
$16.9\times16.9$ arcmin. In SOXS, the chandra\_acisi\_cy22 configuration corresponds to a square field of view of $\simeq 20.008$ arcmin; this slightly larger nominal FoV only affects the empty outer regions of the synthetic images and has no impact on our BCG-offset measurements, which lie well within the ACIS-I array.
The simulated X-ray images are produced in the soft band, restricting photon energies to 0.1–2.0 keV.

\begin{figure*}
    \centering
    \includegraphics[width=\textwidth]{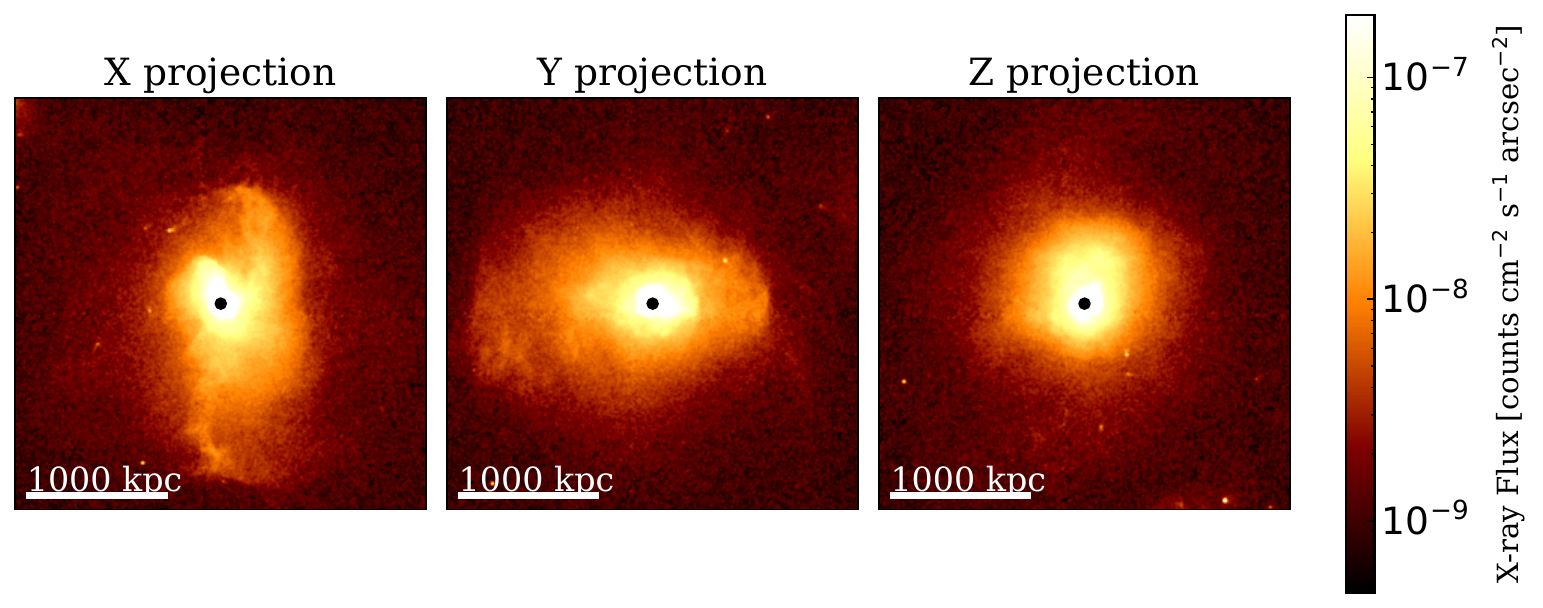}
    \caption{
    Mock X-ray images of FOF halo 0 from three projections (X, Y, Z), centered on the group position and smoothed to a $\sim$10 kpc resolution. Black dots mark the centers used for cropping. A 1000 kpc scale bar is shown in each panel. The color indicates the photon counts, normalized using logarithmic scaling.
    }
    \label{fig:xray_proj}
\end{figure*}

We center the images on the group positions of each halo and generate X-ray photons within a cubic region of side length $2 \times$ Group\_R\_Crit500. To locate the brightest X-ray peaks, we smooth the photon maps with a Gaussian kernel, achieving a final resolution of approximately 10 kpc. Following \citet{prunier2024xraycavitiestngclusteragn}, we define the BCG offset as the projected distance between the brightest X-ray pixel and the position of the most massive black hole within the corresponding FOF halo. Fig.~\ref{fig:xray_proj} shows the X-ray images of FOF halo 0, for which Group\_R\_Crit500 is 1474 kpc.

\subsection{Aperture mass maps simulation}

We develop a fast method for estimating mass ratios using aperture mass maps, validated on both TNG300-1 and LoVoCCS data. In TNG300-1, we select halos at redshift $z = 0.08$ (snapshot 92), chosen to match the typical redshift of the LoVoCCS cluster sample. Our sample includes halos with $M_{200c} > 5\times10^{13}\,M_\odot$ and halos with $10^{13}\,M_\odot < M_{200c} < 5\times10^{13}\,M_\odot$ that have mass ratios greater than 0.1. Here, $M_{200c}$ is defined by the Group\_M\_Crit200 field in the group catalogs as the mass enclosed within a sphere whose mean density is 200 times the critical density of the universe at the halo redshift. Particles within $R_{200c}$ are extracted and projected along the $z$-axis to generate surface density maps. The projection depth along the line of sight is set to $4 \times R_{200c}$.

To simulate the aperture mass signal-to-noise (S/N) map, we begin with the convergence $\kappa(\boldsymbol{\theta})$. After converting $\kappa(\boldsymbol{\theta})$ to the shear $\gamma(\boldsymbol{\theta})$, we combine the resulting shear field with the intrinsic ellipticities of galaxies to generate mock observed ellipticities. The aperture mass maps are then constructed following the LoVoCCS pipeline \citep{fu2024lovoccsiiweaklensing}.

The convergence is defined as the dimensionless surface mass density,
\begin{equation}
\kappa(\boldsymbol{\theta}) = \frac{\Sigma(D_d \boldsymbol{\theta})}{\Sigma_{\rm cr}}, 
\quad 
\Sigma_{\rm cr} = \frac{c^2}{4 \pi G} \frac{D_s}{D_d D_{ds}} \, .
\label{eq:kappa_def}
\end{equation}
Here $\Sigma_{\rm cr}$, $D_d$, $D_s$, and $D_{ds}$ denote the critical density, the angular diameter distances to the lens, to the source, and from the lens to the source, respectively. In our setup, the lens (i.e., the target FOF halos) is placed at redshift 0.08, and the source (i.e., background) is assumed to lie at redshift 1.0.

The Fourier transform of the convergence field $\kappa(\boldsymbol{\theta})$ is defined as
\begin{equation}
\hat{\kappa}(\boldsymbol{\ell}) = \int_{\mathbb{R}^2} d^2\theta \,
\kappa(\boldsymbol{\theta}) \, \exp\!\big(i \boldsymbol{\ell}\cdot \boldsymbol{\theta}\big).
\end{equation}
Following \citet{weaklensing_Bartelmann_2001}, the shear in Fourier space is related to the convergence via
\begin{equation}
\hat{\gamma}(\boldsymbol{\ell}) 
= \hat{\mathcal{D}}(\boldsymbol{\ell})\hat{\kappa}(\boldsymbol{\ell}),
\qquad \boldsymbol{\ell} \neq \boldsymbol{0},
\end{equation}
with the kernel
\begin{equation}
\hat{\mathcal{D}}(\boldsymbol{\ell}) 
= \frac{\ell_1^2 - \ell_2^2 + 2 i \ell_1 \ell_2}{|\boldsymbol{\ell}|^2}.
\end{equation}
The resulting shear field is then transformed back to real space. 

We assume a background galaxy number density of 20 arcmin$^{-2}$ and an intrinsic shape noise of $\sigma_\epsilon = 0.3$, according to \cite{lsstbkgnoises}, noting that the morphology of S/N maps is not strongly sensitive to the choice of $\sigma_\epsilon$. Background galaxies are randomly distributed, and their observed ellipticities are calculated from the shear $\gamma(\boldsymbol{\theta})$ and the convergence $\kappa(\boldsymbol{\theta})$. In the weak lensing limit, where $\kappa \ll 1, \gamma \ll 1$, the observed ellipticity $\epsilon$ can be modeled as
\begin{equation}
\epsilon = 
\displaystyle \frac{\epsilon_{gal} + g}{1 + g^* \epsilon_{gal}}
\end{equation}
where 
\begin{equation}
g(\boldsymbol{\theta}) \equiv \frac{\gamma(\boldsymbol{\theta})}{1 -\kappa(\boldsymbol{\theta})} \, .
\end{equation}
Here, $\epsilon_{gal}$ denotes the intrinsic (unlensed) ellipticities of galaxies. A full derivation is provided in \citet{weaklensing_Bartelmann_2001}.

Based on the observed ellipticities, we compute aperture mass statistics following the LoVoCCS pipeline \citep{LoVoCCSI}, applying the Schirmer filter \citep{Schirmer2004, Schirmer2004etal, Hetterscheidt2005} as the weighting function $Q$ to integrate the tangential shear in Eq. \ref{Q_func}, to match the weighting function used in LoVoCCS, where $x = R/R_{\rm ap}$ with $R_{\rm ap}$ the aperture radius.

The resulting aperture mass S/N map is computed using Equation~\ref{S/N_eq}, with $Q$ as the Schirmer filter, $|\epsilon|$ as the modulus of the complex ellipticity, $\epsilon_i$ the tangential component of the member galaxy ellipticity serving as an estimator of the tangential shear $\gamma_t$ in the weak-lensing limit, and the summation taken over galaxies within the aperture radius.

\begin{equation}
\label{Q_func}
Q(x) = 
\frac{1}{1 + \exp(6 - 150x) + \exp(50x - 47)} 
 \frac{\tanh(x/0.15)}{x/0.15}
\end{equation}

\begin{equation}
\label{S/N_eq} 
\frac{S_i}{N}(\mathbf{R}) = 
\frac{
\sqrt{2} \sum_{\mathbf{R}'} Q(|\mathbf{R}' - \mathbf{R}| / R_{\mathrm{ap}}) \, \epsilon_i(\mathbf{R}', \mathbf{R})
}{
\sqrt{ \sum_{\mathbf{R}'} Q^2(|\mathbf{R}' - \mathbf{R}| / R_{\mathrm{ap}}) \, |\epsilon(\mathbf{R}')|^2 }
}
\end{equation}

\begin{figure*}
    \centering
    \includegraphics[width=\textwidth]{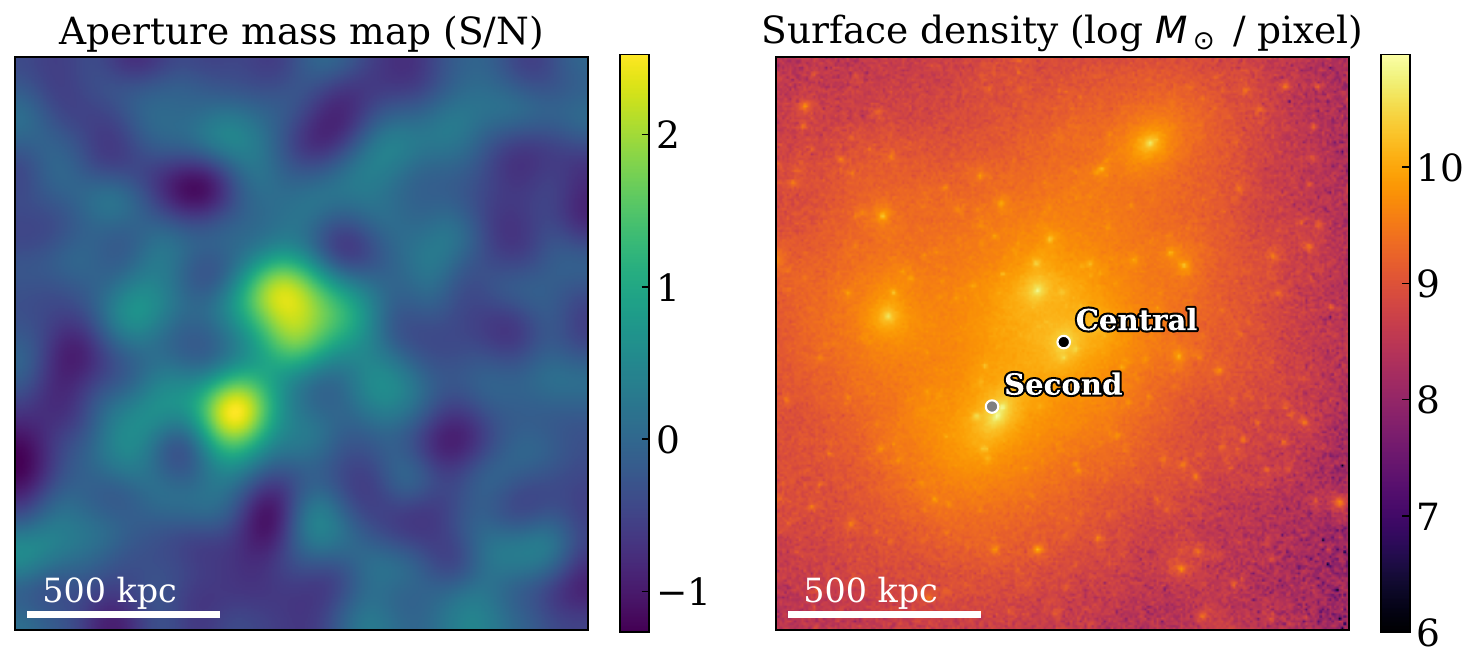}
    \caption{
        Example mock observation from the TNG300-1 simulation. A 500 kpc scale bar is shown in each panel.
        Left: Simulated aperture mass map (S/N) of FOF halo 142, with $R_{200c} \sim 750$kpc, incorporating shape noise consistent with observational weak lensing data. For visualization, the maps are Gaussian-smoothed with a kernel scale of $\sim$ 60 kpc; this smoothing is applied for display purposes only.
        Right: Surface mass density map (dark matter + gas) in units of $\log\,M_\odot$/pixel. The positions of the central and the most massive satellite subhalos are marked for reference.
    }
    \label{fig:mock_aperture_mass}
\end{figure*}

As halos in TNG300-1 are less massive than LoVoCCS clusters, to enhance the signal-to-noise ratio for low-mass halos, we scale the background galaxy density by a factor of $(M_{\rm halo}/5 \times 10^{14} M_\odot)^2$, where $5 \times 10^{14} M_\odot$ is approximately the median mass of the LoVoCCS clusters. We define the mass ratio of each FOF halo as the mass of the most massive satellite subhalo (within the region used to generate the aperture mass maps) divided by the mass of the central subhalo. Fig.~\ref{fig:mock_aperture_mass} shows an example of our mock observations for FOF halo 142 in the TNG300-1 simulation.

\subsection{Mass ratio estimation}
We use the peak\_local\_max function from scikit-image \citep{scikit_image_van_der_Walt_2014} to identify the brightest peaks in the aperture mass maps generated from both TNG300-1 and LoVoCCS clusters. Each map covers a field of view of $2R_{200}$ on a side and is discretized on a $240\times240$ grid, corresponding to a pixel size of $\Delta x = (2R_{200})/240 = R_{200}/120$ in physical units. After applying a ten-pixel Gaussian smoothing, the two highest peaks are assumed to correspond to the central and the most massive satellite subhalos. We then construct an empirical quantity, defined in Equation~\ref{eq: rough_quantity}, by taking the ratio of their peak intensities and multiplying it by their separation distance. By correlating this constructed quantity with the true mass ratios, we aim to establish a proxy that can be applied to observations with known mass ratios.

In the TNG300-1 simulation, the true mass ratio is defined as the ratio of the SubhaloMass values in the group catalog, which represents the total bound mass of each subhalo. In the LoVoCCS clusters, we estimate the mass ratio using weak lensing reconstruction pipelines, under the assumption that the two peaks correspond to two distinct cluster centers.

\begin{equation}
    q = \frac{I_2}{I_1} \times R_{\mathrm{sep}}
    \label{eq: rough_quantity}
\end{equation}
where $I_1$ and $I_2$ are the intensities of the first and second brightest peaks in the aperture mass map, and $R_{\mathrm{sep}}$ is their projected separation on the map. In practice, we measure $R_{\rm sep}$ in pixel units on the $240\times240$ grid; the corresponding physical separation is $R_{\rm sep}\,\Delta x$, where $\Delta x=(2R_{200})/240$ is the map pixel size. Since $I_2/I_1$ is dimensionless, $q$ has the same units as $R_{\rm sep}$. We note that this empirical proxy may be sensitive to the choice of smoothing kernel and the relative sizes of the cluster and its subhalos. In particular, an overly small kernel may underestimate the peak intensity of extended subhalos, while a large kernel can smear out more compact structures. The optimal kernel size is non-trivial to determine and may vary across systems; in this work, we adopt a fixed value because our aim is to demonstrate correlation rather than optimize the mass fraction measurement.

\section{Results}
\subsection{Phase-space disturbance}
\subsubsection{Decay timescale selection}
\label{sec: Decay Timescale Selection}
To quantify the connection between dynamical disturbance and merger history, we first determine a proper decay timescale $\tau$ in the merger score definition (Eq.~\ref{eq:merger_score}). This parameter controls how strongly recent mergers contribute relative to older ones. For this pre-evaluation, we include all candidate features and will prune them after selecting a suitable $\tau$. We evaluate the model performance based on $R^2$ in the training and test datasets. Here $R^2$ denotes the standard coefficient of determination computed between the true disturbance scores $s_i$ and the predicted scores $\hat{s}_i$,
\begin{equation}
R^2 \equiv 1 - \frac{\sum_i (s_i-\hat{s}_i)^2}{\sum_i (s_i-\bar{s})^2},
\end{equation} 
where $\bar{s}$ is the mean of the true scores. Thus, $R^2=1$ corresponds to a perfect prediction, $R^2=0$ indicates performance no better than predicting the mean, and negative values imply worse-than-baseline performance.

The choice of $\tau$ is guided by two considerations. First, we evaluate model performance by training XGBoost \citep{Chen_2016_XGBoost}, via scikit-learn \citep{pedregosa2018scikitlearnmachinelearningpython} to predict full-, past-, and future-merger scores from all phase-space features, and tracking how the test $R^2$ varies with $\tau$. Second, we account for sample availability: halos within $\tau$ Gyr of redshift 0 lack complete future-merger information and must be excluded when using full- or future-merger scores. Increasing $\tau$ enlarges this exclusion window and reduces the effective sample size. XGBoost, a scalable tree boosting system, is chosen for its ability to capture nonlinear relationships and its interpretability via feature-importance metrics.

To increase both sample size and model robustness, we treat the three projections of each of the 352 primary zoom-in targets and their progenitors between snapshots 72 and 99 as distinct samples. When splitting the dataset into training and test sets, we use grouping: we assign each sample a group label and perform the split at the level of group labels, i.e., all samples sharing the same label are placed entirely in either the training set or the test set. However, the degree of data leakage can depend on the chosen grouping scheme. Grouping by both halo ID and snapshot number can prevent leakage between different projections of the same halo at the same snapshot, but it still allows different snapshots of the same halo to fall into different splits and can therefore introduce temporal leakage since adjacent snapshots are strongly correlated, e.g. the model might learn to predict the merger score at snapshot 73 by implicitly memorizing the features of the same halo at snapshot 72.

To mitigate this, we apply a projection-rotation scheme along with snapshot dropping. Specifically, we alternate between using different phase-space projections (x, y, z) and exclude the snapshots in between. For example, in the sequence [z, drop, x, drop, y, drop], each included snapshot has a distinct projection and is separated from its adjacent neighbors. This setup helps reduce the risk of data leakage by discouraging the model from learning patterns that depend on the similarity of halo structures across consecutive snapshots, which may act as a proxy for halo identity.

To assess potential data leakage, we conduct two complementary tests. First, we experiment with a more conservative grouping strategy where we split data only by halo ID, without including snapshot information. This setup leads to a slightly lower test $R^2 \sim 0.40$ and training $R^2 \sim 0.70$ for past-merger scores at $\tau=2.0$ Gyr, as described in \ref{sec: Data Leakage-Minimized Baseline}, compared to the results grouped by both halo ID and snapshot number ($R^2_{\rm test} \sim 0.55$, $R^2_{\rm train} \sim 0.80$). This comparison raises concern about possible data leakage when grouping by both halo ID and snapshot number. To further test for potential data leakage, we construct a dedicated test set by randomly selecting 20\% of halos and excluding them entirely from the training process. We then apply the same training configuration as in the main analysis, including projection-rotation and grouping by both halo ID and snapshot, and evaluate model performance on both the internal test split and the pre-separated test sets. The model achieves consistent scores across both sets. These two additional tests might imply that our projection-rotation scheme and grouping strategy do not lead to severe data leakage while increasing sample diversity. 

We note that the snapshots are spaced in redshift rather than in equal time intervals, so halos near redshift 0.4 are separated by shorter physical times. As a result, their structural properties change little across snapshots, increasing the risk that the model learns to associate nearly identical halos across time steps rather than extracting generalizable features. Furthermore, for large values of $\tau$, merger scores become more similar across nearby snapshots, raising the chance that the model relies on learning from a halo’s immediate progenitors or descendants.

\begin{figure*}
    \centering
    \includegraphics[width=\textwidth]{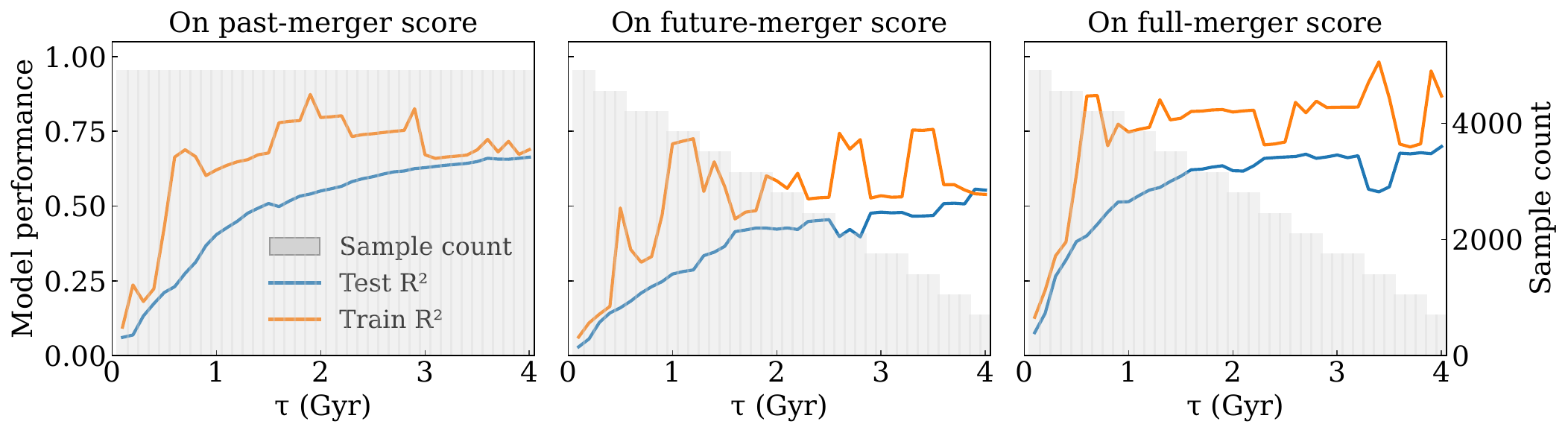}
    \caption{
        Model performance as a function of $\tau$, evaluated using three target score definitions: past-merger (left), future-merger (middle), and full-merger (right). 
        Orange lines show the $R^2$ on the training set, and blue lines show the $R^2$ on the test set. 
        All results are based on the rotated projection configuration described in Section \ref{sec: Decay Timescale Selection}.
    }
    \label{fig:model_tau_scan}
\end{figure*}

We perform a grid search for each $\tau$, using learning rates in {0.005, 0.01, 0.05, 0.1, 0.2}, maximum tree depths in {1, 2, 4, 6, 8, 10}, and numbers of estimators in {100, 200, 300, 500, 1000}. Each model is trained with its own optimized hyperparameters under this scheme, and the resulting performance is used to compare across different $\tau$ values. The scan results are shown in Fig.~\ref{fig:model_tau_scan}.

Considering both the limited sample size and the risk of data leakage, we adopt a small value of $\tau = 2.0$ Gyr, at which the training and test $R^2$ curves begin to stabilize. To ensure a fair comparison across the three merger score definitions, we match their time windows: the future- and past-merger scores are computed over a window of $\tau$, while the full-merger score uses a symmetric window of $\tau/2$ before and after the snapshot. As shown in Fig.~\ref{fig:tau_efficiency}, the model trained on past-merger scores performs slightly better. However, this advantage may be partly due to sample loss in the other two definitions at larger $\tau$. Overall, the differences in model performance are minor, suggesting that our phase-space features alone are not sufficient to distinguish between different merger stages. The input features are summarized in the right panel of Fig.~\ref{fig:tau_efficiency}. We define $\Delta\mathrm{bic}$ as $\mathrm{bic}2 - \mathrm{bic}1 - 6\ln(n)$, which is independent of $n$, the number of member galaxies. To see why this definition is independent of richness, recall that for a $K$-component 2D GMM, BIC can be computed via Eq. \ref{eq:bic_eq}. For our 2D GMMs, $k_1=5$ and $k_2=11$, hence $\Delta\mathrm{bic}\equiv(\mathrm{bic}2-\mathrm{bic}1)-6\ln N$ removes the explicit $\ln N$ term and is therefore independent of richness. The method for calculating $\lambda_2/\lambda_1$ is described in Section \ref{sec: Phase-space Feature Extraction}.

\begin{figure*}[t]
\centering
\begin{minipage}[t]{0.56\textwidth}
    \vspace*{0pt} 
    \centering
    \includegraphics[width=\textwidth]{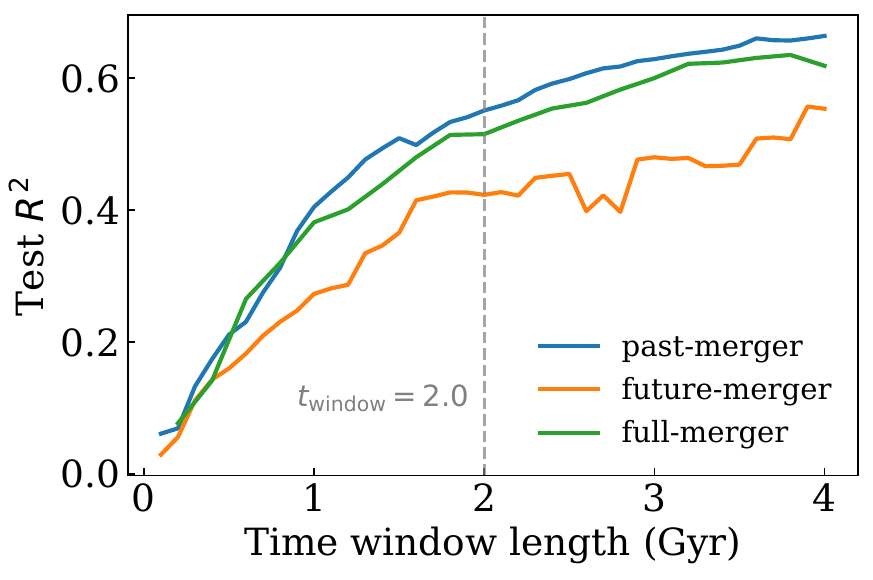}
    \label{fig:tau_efficiency}
\end{minipage}%
\hfill
\begin{minipage}[t]{0.42\textwidth}
    \vspace*{0pt} 
    \centering
    \footnotesize
    Summary of input features used in the model.\\[3pt]
    \begin{tabular}{ll}
    \hline
    Feature & Description \\
    \hline
    $\Delta r$ & Projected separation of two GMM groups [kpc] \\
    $v_1$, $v_2$ & Mean velocities of groups 1 and 2 [km/s] \\
    $\Delta v$ & Velocity difference between groups [km/s] \\
    $\sigma_{r,1}, \sigma_{r,2}$ & Projected radius dispersions [kpc] \\
    $\sigma_{v,1}, \sigma_{v,2}$ & Velocity dispersions [km/s] \\
    $f_n$ & Number fraction of the smaller group \\
    bic1, bic2 & BIC from 1- and 2-component GMM fits \\
    $\Delta$bic & Penalized BIC difference \\
    $\lambda_2/\lambda_1$ & Ellipticity (elongation ratio) \\
    $z$ & Redshift \\
    $f_m$ & Mass ratio \\
    $\sigma_r$, $\sigma_v$ & Global radius dispersion [kpc], \\
                          & and velocity dispersion [km/s] \\
    \hline
    \end{tabular}
    \label{table:feature}
\end{minipage}

\vspace{6pt}
\caption{
Test $R^2$ as a function of the time window $\tau$ for the past-, future-, and full-merger scores, under matched time window conditions. 
The right panel summarizes the phase-space input features used in the model.
}
\label{fig:tau_efficiency}
\end{figure*}

\subsubsection{Model performance at selected $\tau$}
\label{sec: Model Performance at Selected tau}
\begin{figure*}
    \centering
    \includegraphics[width=\textwidth]{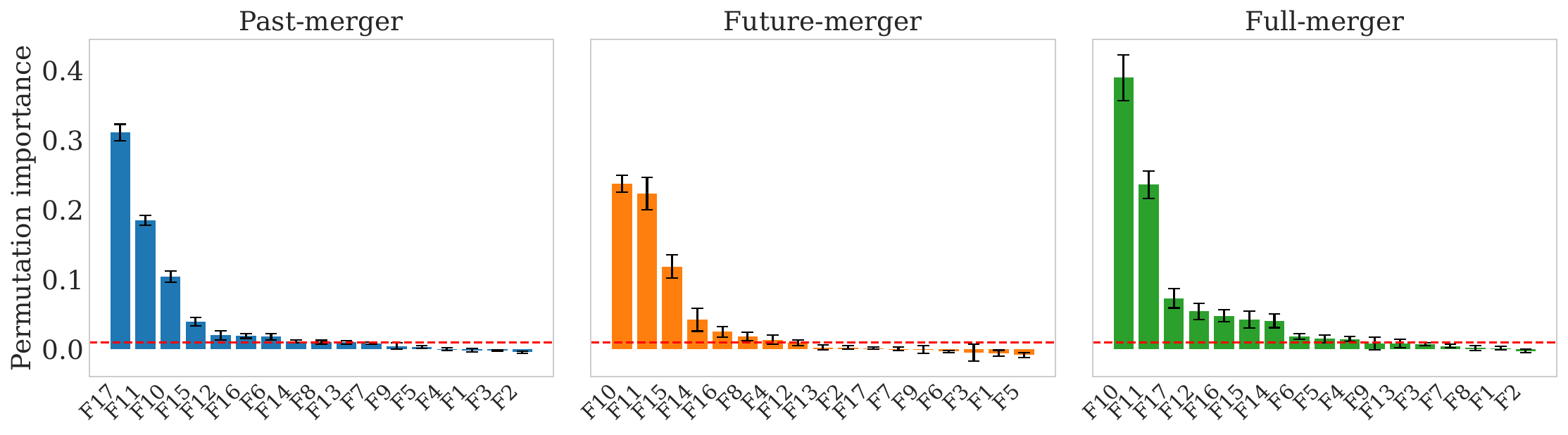}
\caption{
Comparison of permutation feature importance rankings for models using the three merger score definitions: past-merger (left), future-merger (middle), and full-merger (right). 
Bars indicate the mean importance of each feature, with error bars representing the standard deviation over 10 independent permutations. 
The red dashed line marks the 0.01 threshold used for feature selection. 
Features are labeled F1–F17, ranked by importance within each model. Their definitions are as follows: 
F1: $\Delta R$, 
F2: $v_1$, 
F3: $v_2$, 
F4: $\Delta v$, 
F5: $\sigma_{r,1}$, 
F6: $\sigma_{r,2}$, 
F7: $\sigma_{v,1}$, 
F8: $\sigma_{v,2}$, 
F9: $f_n$, 
F10: bic1, 
F11: bic2, 
F12: $\Delta\mathrm{bic}$, 
F13: $\lambda_2/\lambda_1$, 
F14: $z$, 
F15: $f_m$, 
F16: $\sigma_r$, 
F17: $\sigma_v$.
}
\label{fig:feature_importance}
\end{figure*}

\begin{figure*}[htbp]
    \centering
    \includegraphics[width=\textwidth]{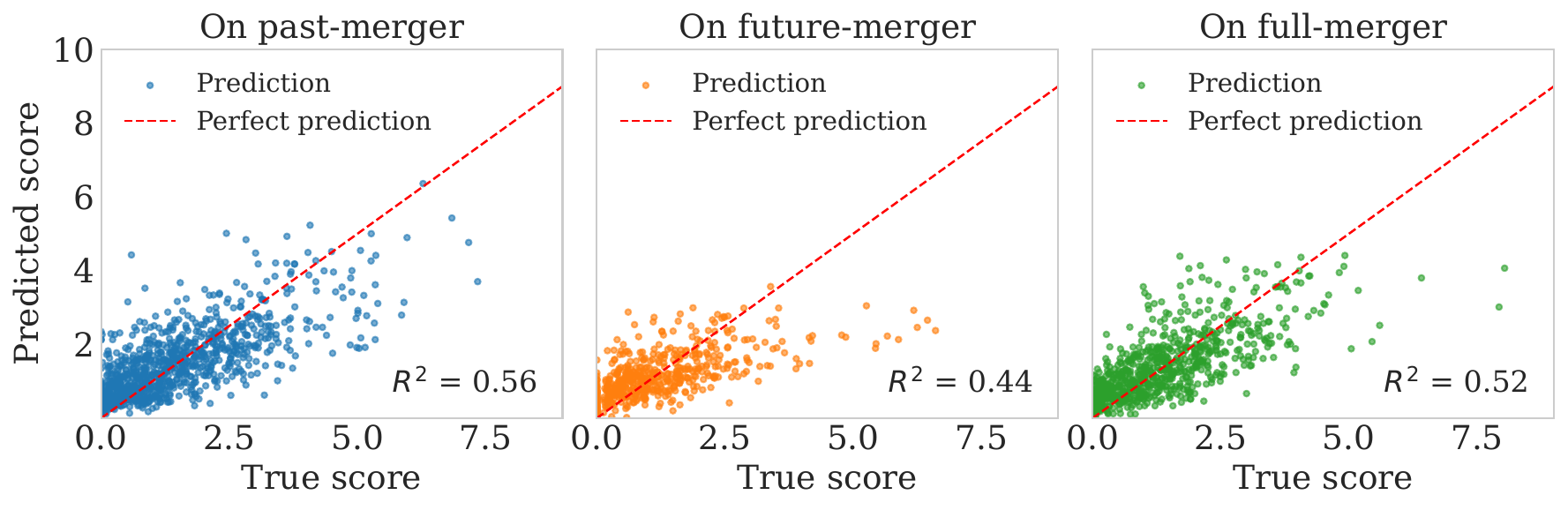}
    \caption{
        Predicted merger scores versus true scores for three test sets: on past-mergers (left), on future-mergers (middle), and on full-merger samples (right). 
        The red dashed line indicates the ideal case of perfect prediction. 
        Each panel also displays the $R^2$ score for the corresponding test. 
        Points are slightly transparent to indicate density. The model shows consistent predictive performance across different temporal subsets. High true scores tend to be underpredicted. This is mainly because the strongly skewed true-score distribution (Fig.~\ref{fig:score_distribution}): most halos have low scores, which dominate the training and bias the predictions toward low values.
    }
    \label{fig:score_scatter}
\end{figure*}

\begin{figure*}[htbp]
    \centering
    \includegraphics[width=\textwidth]{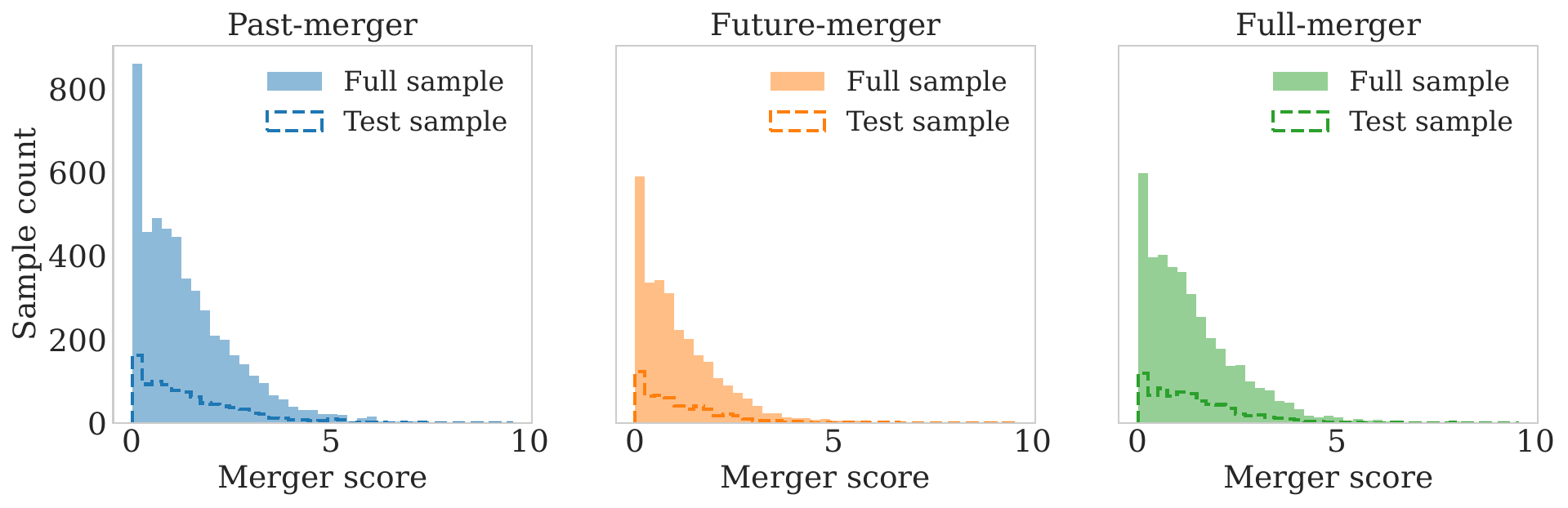}
\caption{
Distribution of true merger scores for the full sample (filled bars) and the test sample (outlined bars), shown for the past-merger, future-merger, and full-merger datasets. 
}
\label{fig:score_distribution}
\end{figure*}

We evaluate permutation feature importance for models trained on the three merger score definitions, using a time window of 2.0 Gyr. Specifically, this corresponds to $\tau = 2.0$ Gyr for the past- and future-merger scores, and a symmetric window of $\tau = 1.0$ Gyr for the full-merger score. Permutation feature importances for all features are presented in Fig.~\ref{fig:feature_importance}, and we retain only those with a relative importance greater than 0.01 in at least one of the three models. The resulting set includes $\Delta v$, $\sigma_{r,1}$, $\sigma_{r,2}$, $\sigma_{v,2}$, bic1, bic2, $\Delta\mathrm{bic}$, $\lambda_2/\lambda_1$, $z$, $f_m$, $\sigma_r$, and $\sigma_v$. Although $\sigma_{v,1}$ does not meet the $>0.01$ criterion, we add it to the pruned feature set for symmetry, given that $\sigma_{v,2}$ is retained and the grouping depends on the adopted GMM fitting method. Fig.~\ref{fig:score_scatter} presents scatter plots of the predicted versus true scores for all three models trained on the pruned feature set, while Fig.~\ref{fig:score_distribution} shows the corresponding score distributions.

\subsection{Photometric differentiation of merger stages}
\label{sec:Photometric Differentiation of Merger Stages}
To investigate whether photometric features can differentiate past from future mergers, we compare results using the past- and future-merger scores defined in Section \ref{sec: Merger Score Definition}. We select 352 primary zoom-in targets for snapshots between redshift 0.4 and 0 (corresponding to snapshot 72 to 99), treating each snapshot of a halo as an independent sample.

We adopt a time window of 2.0 Gyr to track the evolution of the SFR, measuring it from 1 Gyr before to 1 Gyr after each snapshot. To ensure reliable curvature estimates, we discard SFR tracks with fewer than three time points, which typically arises from limited redshift resolution or from the restricted redshift range of the analysis. While 2.0 Gyr is somewhat longer than the characteristic timescale of major merger events, a shorter window would reduce the number of available samples due to resolution limits and lead to larger uncertainties. We also tested a shorter window of 1.0 Gyr and found that the results remain consistent.

For correlations involving the future-merger score, we further exclude samples within 2.0 Gyr of redshift 0, where future merger activity cannot be reliably traced. The curvature of normalized SFR is computed as described in Section \ref{sec: Star Formation and Blue Galaxy Fraction Statistics}.

\begin{figure*}[htbp]
\sidecaption
    \includegraphics[width=12cm]{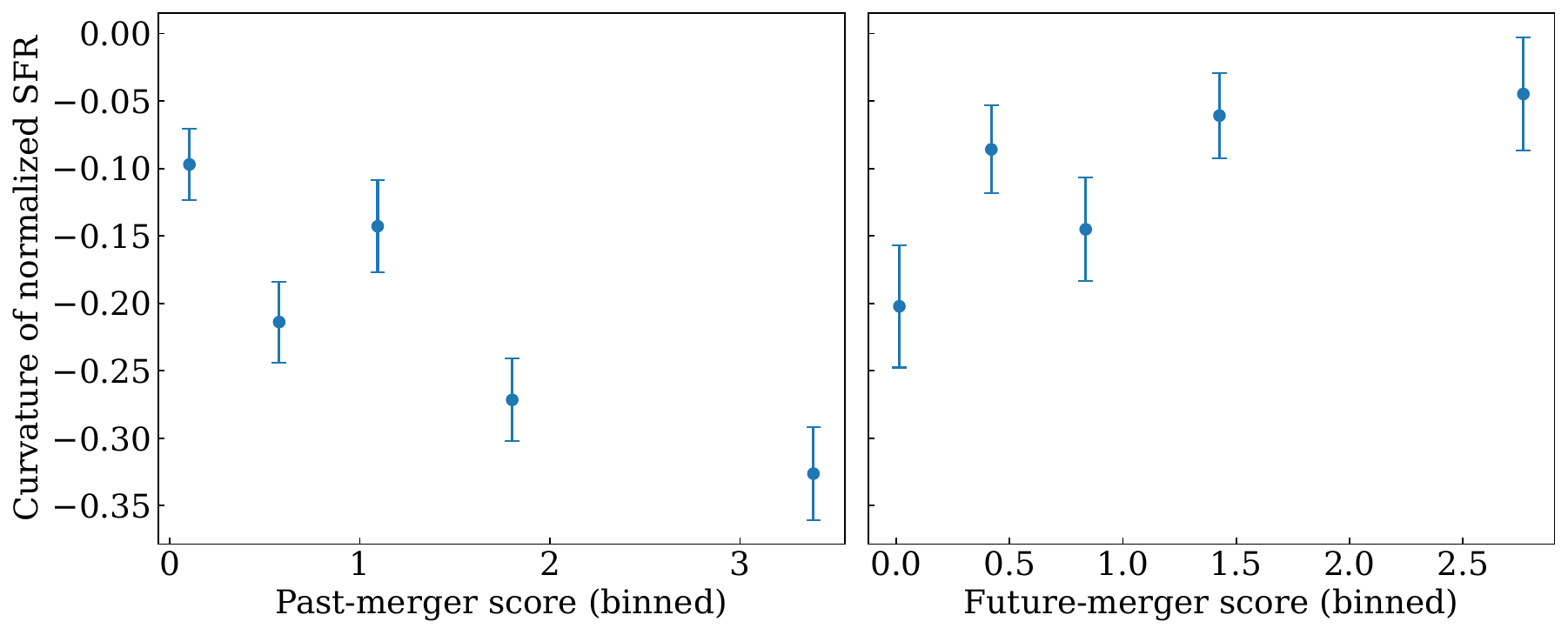}
    \caption{
        Curvature of the normalized SFR as a function of merger score.
        The left and right panels show results for the past- and future-merger scores, respectively.
        In each case, the score distribution is divided into five quantile bins.
        A total of 9812 samples are used in the past-merger analysis and 5609 in the future-merger case.
        Error bars indicate the standard error of the mean curvature within each bin.
    }
    \label{fig:sfr_curvature_merger}
\end{figure*}

As shown in Fig.~\ref{fig:sfr_curvature_merger}, the correlations between SFR curvature and merger scores exhibit distinct trends for past and future mergers.

\begin{figure*}
    \sidecaption
    \includegraphics[width=12cm]{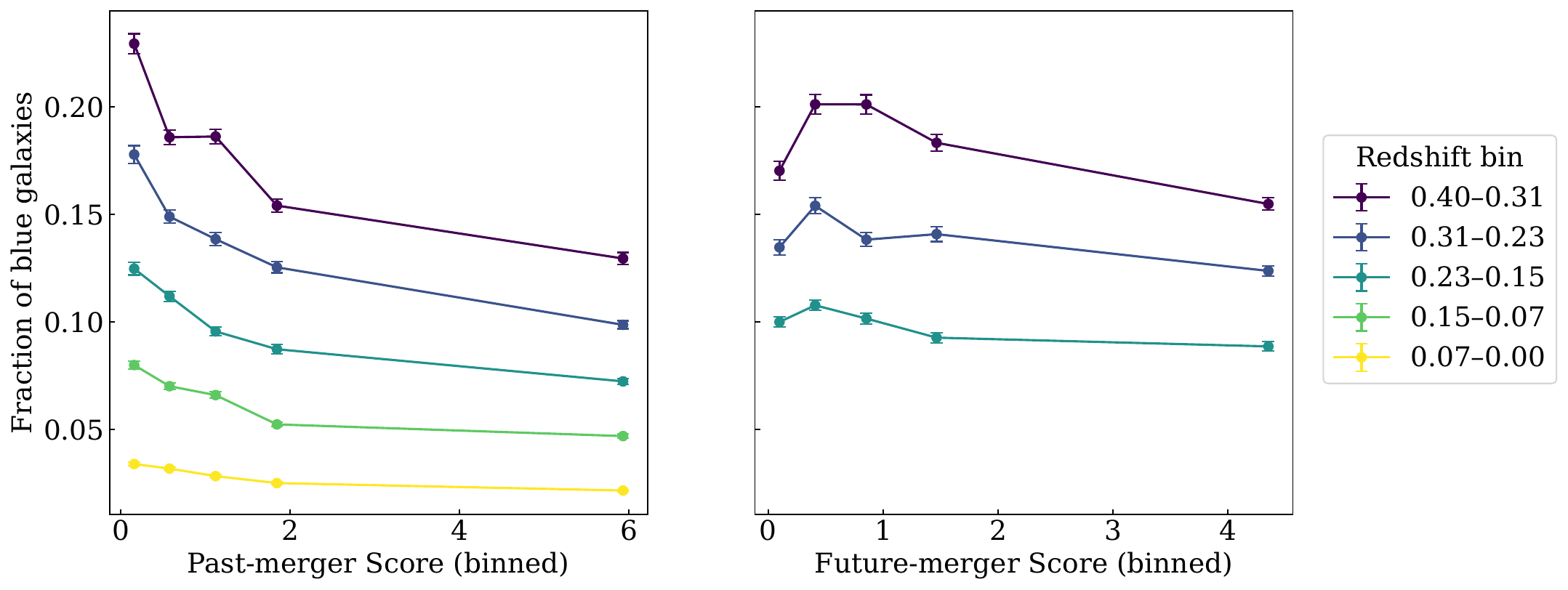}
    \caption{
    Fraction of blue galaxies as a function of merger score, shown separately for past-merger scores (left) and future-merger scores (right). 
    Each curve corresponds to a redshift bin defined as $[z_{\mathrm{low}}, z_{\mathrm{high}})$, i.e., including the lower bound but excluding the upper bound. 
    Within each redshift bin, the score distribution is divided into five quantile bins.
    Error bars indicate the standard error of the mean blue fraction within each merger score bin.
    Several redshift bins are absent in the future-merger panel due to the exclusion of low-redshift samples ($z < 0.15$).
    }
    \label{fig:blue_frac_vs_score}
\end{figure*}

Motivated by the SFR results, we investigate whether a directly observable quantity can serve as a proxy for distinguishing between the two merger scores, a task for which phase-space-based machine learning methods have shown limited success, as shown in Section \ref{sec: Decay Timescale Selection}. We compute the blue galaxy fraction as described in Section \ref{sec: Star Formation and Blue Galaxy Fraction Statistics}, and exclude samples below redshift 0.15 in the future-merger analysis to ensure sufficient look-forward time. As shown in Fig. \ref{fig:blue_frac_vs_score}, the trends differ noticeably between the two cases, suggesting that the blue galaxy fraction may carry physical information relevant to merger stages.

\subsection{BCG offset as a morphological merger indicator}

We exclude a small number of clusters with BCG offsets exceeding 500 kpc, as such large displacements are not detectable in observations and primarily result from definitional artifacts in the simulation. Specifically, our BCG offset is defined as the distance between the X-ray peak and the most massive black hole, which can reside in different progenitors during major mergers.

\begin{figure*}[htbp]
    \centering
    \includegraphics[width=\textwidth]{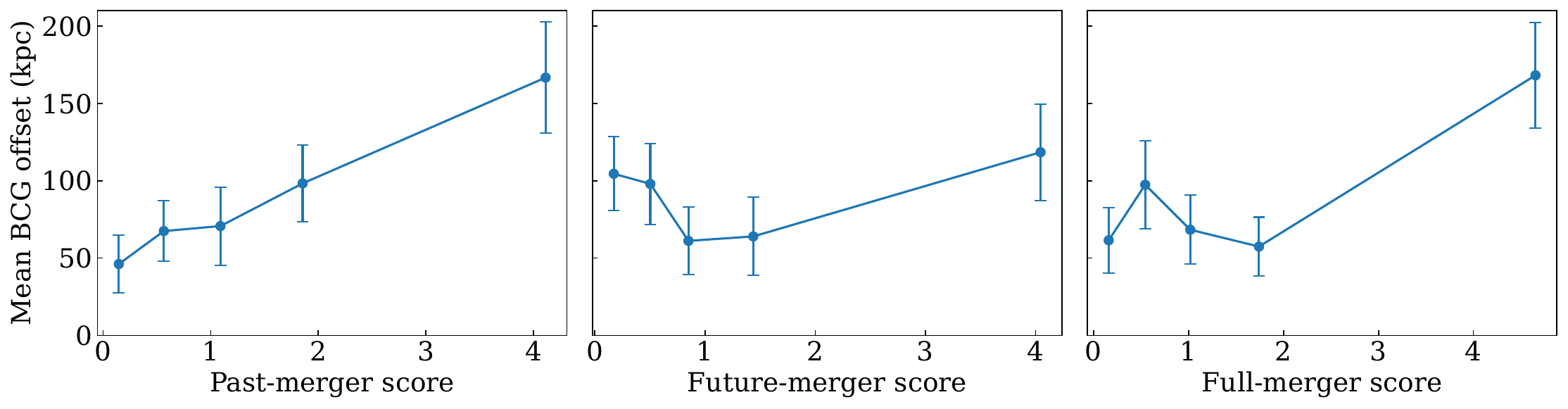}
    \caption{
    Mean BCG offset as a function of merger score under three definitions: past-merger (left), future-merger (middle), and full-merger (right). Each point represents the average offset within a merger score bin, with vertical bars indicating the standard error of the mean. The total sample size is 325 clusters. To ensure roughly equal sample counts in each bin, we apply quantile-based binning: full- and past-merger scores are divided at $[0.0, 0.2, 0.4, 0.6, 0.8, 1.0]$; future-merger scores are divided at $[0.0, 0.3, 0.48, 0.66, 0.84, 1.0]$ due to the heavy concentration near zero. A clear positive trend is visible in the first panel, suggesting that BCG displacement increases with merger activity based on past events.
    }
    \label{fig:bcg_vs_merger_score}
\end{figure*}

As shown in Fig.~\ref{fig:bcg_vs_merger_score}, the BCG offset measured along the $x$-projection shows a strong positive correlation with the past-merger score, suggesting that our catalog-based merger scores are consistent with traditional dynamical disturbance indicators. Notably, BCG offsets are more sensitive to past mergers than to upcoming ones, revealing a time asymmetry that phase-space features do not capture. To ensure a fair comparison across different merger score definitions, we adopt matched time windows: $\tau = 2.0$ Gyr for both the past- and future-merger scores, and a symmetric window of $\tau = 1.0$ Gyr for the full-merger score.

Since BCG offsets measured along different projection axes are highly correlated, with Pearson correlation coefficients of $r = 0.85$, $0.89$, and $0.83$ for the $xy$, $xz$, and $yz$ projection pairs, respectively, we use only the $x$-projection in our analysis.

\subsection{Mass ratio estimation}
\label{sec: Mass Ratio Estimation}

We validate our mass ratio estimator using the LoVoCCS cluster sample by correlating the derived peak intensity ratio (see Eq.~\ref{eq: rough_quantity}) with the true mass ratios. The true mass ratio is inferred from the global two–component weak–lensing mass model: we assume that the two subclusters are located at the brightest and second–brightest peaks in the aperture–mass map, and we use the LoVoCCS mass–fitting pipeline to derive the masses of these two components, as described in \cite{McCleary_2015}. The ratio of these fitted masses is used as the true mass ratio for the observed clusters. After excluding halos with identified mass ratios greater than one, caused by incorrect peak identification (e.g., misidentification of the central or secondary subhalos), we find a Pearson correlation coefficient of $r = 0.31$.

\begin{figure*}
\sidecaption
    \includegraphics[width=12cm]{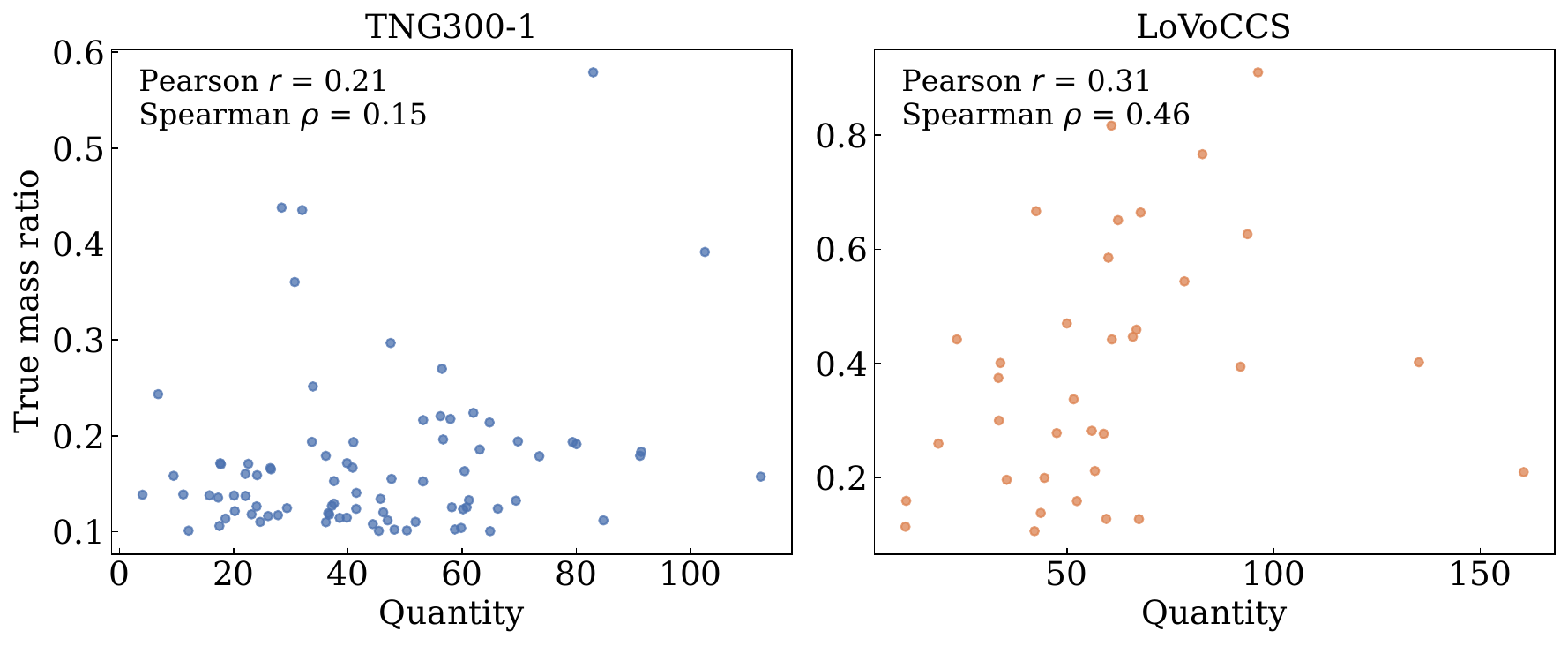}
    \caption{
        Correlation between Quantity, i.e. $q$ in Equation \ref{eq: rough_quantity}, and true merger mass ratio.
        Left: Results from TNG300-1 halos.
        Right: LoVoCCS observed clusters.
        Each point represents a galaxy cluster; the $x$-axis shows the quantity estimator defined in Eq.~\ref{eq: rough_quantity}, and the $y$-axis shows the true mass ratio.
        We report both Pearson $r$ and Spearman $\rho$ correlation coefficients to characterize the relationship.
    }
    \label{fig:mass_ratio_proxy}
\end{figure*}

We further test the estimator with TNG300-1 halos, selecting those with $M_{200c} > 5\times10^{13} M_\odot$ as well as halos with $10^{13} M_\odot<M_{200c} < 5\times10^{13} M_\odot$ that have true mass ratios above 0.1. Applying the same method using a 10-pixel Gaussian smoothing scale on the aperture mass maps ($240\times240$ pixels), consistent with the LoVoCCS setup, yields a Pearson correlation coefficient of $r = 0.49$. Results for both LoVoCCS and TNG300-1 are shown in Fig.~\ref{fig:mass_ratio_proxy}. In Fig.~\ref{fig:mass_ratio_proxy}, the mass ratios in the LoVoCCS clusters are typically larger than those in TNG300-1. This likely reflects a difference in how the mass ratio is defined. In TNG300-1 we are looking at the scale of a single host cluster, so the mass ratio is defined as that between the most massive satellite galaxy and the central subhalo. In LoVoCCS, however, the field of view is typically larger, so the fitted masses can instead correspond to two nearby cluster-scale halos. This difference does not strongly affect our main goal, which is to build a mass-ratio estimator based on aperture-mass maps of two dominant matter clumps, regardless of whether they correspond to a cluster–subcluster pair or to two neighboring clusters. The relatively low correlation coefficient between the true mass ratio and our estimator is expected for several reasons. First, most of the clusters in our sample have small mass ratios (e.g. $\leq$ 0.3), so that even modest observational noise or peak–finding uncertainties translate into large relative errors on the inferred ratio. Second, our estimator is based on identifying and measuring the two brightest peaks in the aperture–mass map. Small misidentifications (e.g., confusing a noise peak) or small shifts in the peak position can significantly change the measured peak ratio. For these reasons we do not expect $q$ to be a precise mass–ratio estimator, but rather a coarse indicator that is primarily useful when combined with other features in the classifier.

\section{Discussion}
\subsection{Data leakage-minimized baseline}
\label{sec: Data Leakage-Minimized Baseline}
\begin{figure*}[htbp]
    \centering
    \includegraphics[width=\textwidth]{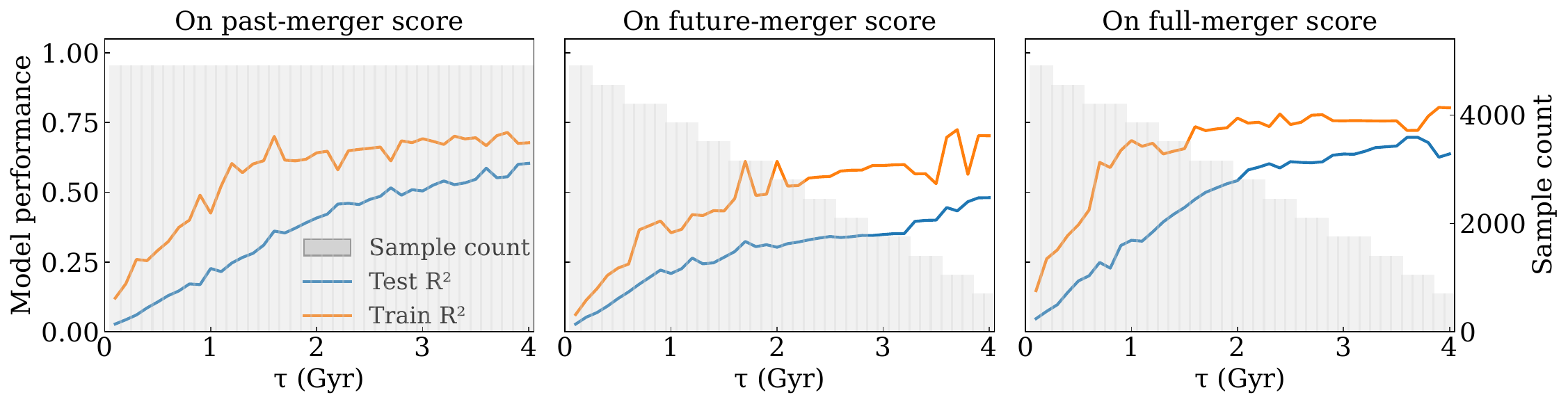}
    \caption{
Model performance as a function of $\tau$, evaluated using three target score definitions: past-merger (left), future-merger (middle), and full-merger (right).
Orange and blue curves show the training and test $R^2$, respectively, and gray histograms indicate the sample count.
This test uses the leakage-minimized split (grouping by halo ID only, so that progenitors/descendants of training halos are excluded from the test set) together with the rotation-based filtering scheme for sample selection.
}
    \label{fig:tau_scan_rotation_baseline}
\end{figure*}

To avoid potential data leakage, we adopt a conservative train-test splitting strategy: for each halo in the training set, none of its progenitors or descendants are included in the test set. This ensures that the model cannot learn merger scores through evolutionary connections. For this evaluation, we use a consistent set of halos selected using the rotation-based filtering scheme described in Section \ref{sec: Decay Timescale Selection}. Models are trained on all candidate features (Fig. \ref{fig:tau_efficiency}) for each of the three merger score definitions, and we scan across values of $\tau$. The hyperparameters are kept consistent with those used in Section \ref{sec: Decay Timescale Selection}. As shown in Fig. \ref{fig:tau_scan_rotation_baseline}, the train and test $R^2$ curves do not differ significantly from those in Fig. \ref{fig:model_tau_scan}, suggesting that no strong data leakage is present.

\subsection{Projection robustness}
\label{sec: Projection Robustness}
\begin{figure*}[htbp]
    \centering
    \includegraphics[width=\textwidth]{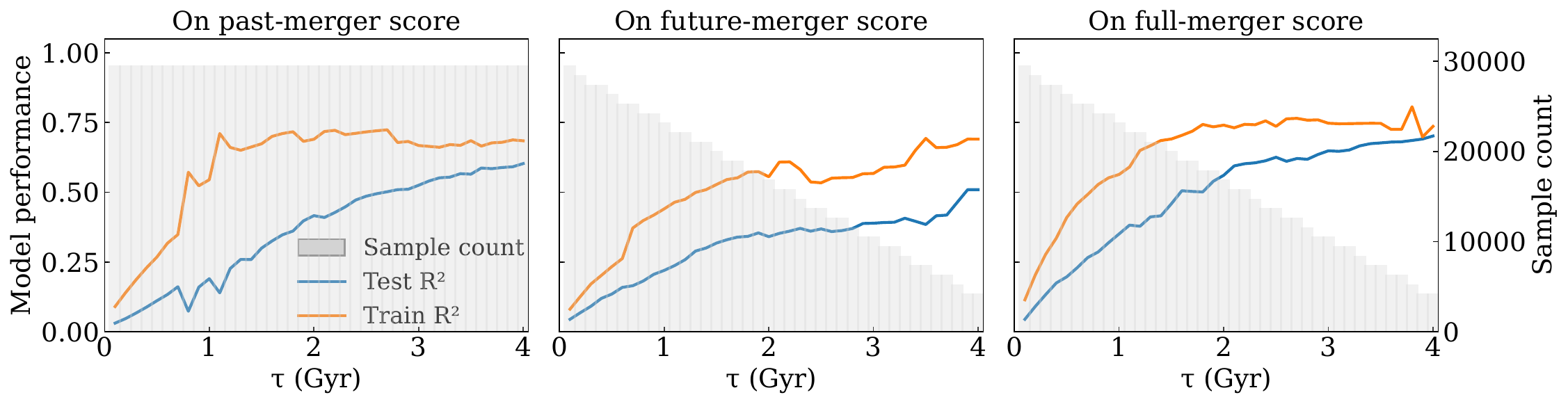}
    \includegraphics[width=\textwidth]{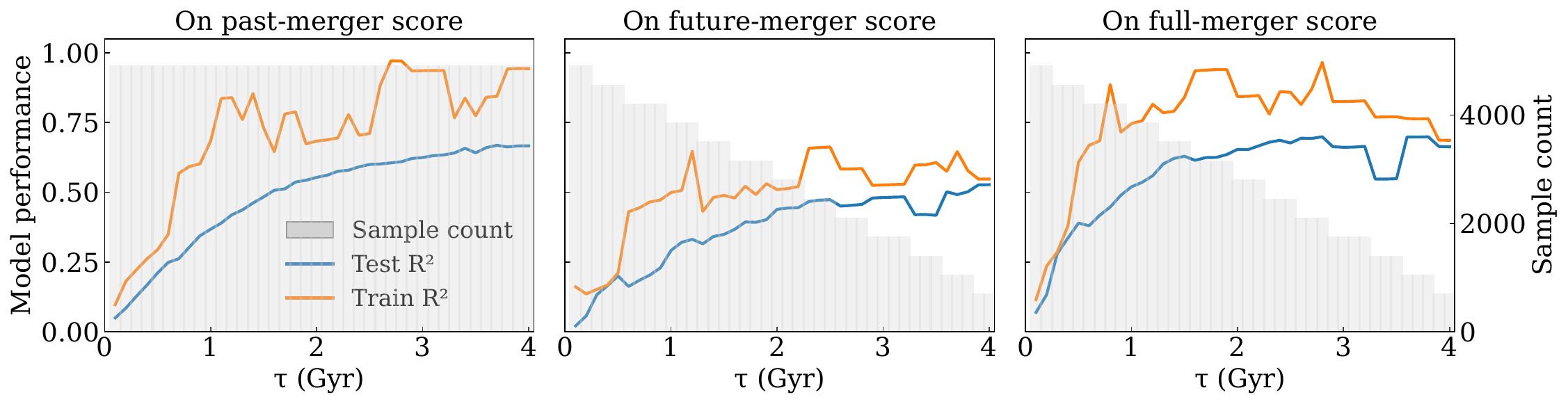}
    \caption{
    Model performance as a function of $\tau$, evaluated using three target-score definitions: past-merger (left), future-merger (middle), and full-merger (right).
    Orange and blue curves show the training and test $R^2$, respectively, and gray histograms indicate the sample count.
    These tests assess projection robustness.
    Top: treating the three orthogonal projections of each halo as independent samples, using the leakage-minimized split (grouping by halo ID).
    Bottom: Following the setup in Section~\ref{sec: Decay Timescale Selection} (grouping by halo ID and snapshot number and using rotation-based filtering), the projection-rotation order is changed from [z, drop, x, drop, y, drop] to [x, drop, y, drop, z, drop].
    The similar performance suggests that the model is not sensitive to the projection order, which is favorable for observational applications where only one projection is available.
    }
    \label{fig:tau_scan_projection}
\end{figure*}

To assess whether additional information can be extracted from the three projections of each FOF halo, we track 352 primary zoom-in targets from redshift 0.4 to 0 and treat the three projections of each halo as independent samples. To minimize data leakage, we apply the same grouping strategy as in Section \ref{sec: Data Leakage-Minimized Baseline}, grouping by halo ID when splitting the training and test sets. The model is trained using all candidate features listed in Fig. \ref{fig:tau_efficiency}, and we scan across different values of $\tau$ for all three merger score definitions, as shown in Fig. \ref{fig:tau_scan_projection}. The performance does not show significant improvements, which might imply that the phase-space diagram generated from a single projection is enough to carry the information we need statistically.

We further test the robustness of our phase-space machine learning model by changing the rotation sequence of projections. While still following the [axis, drop, axis, drop, axis, drop] pattern described in Section \ref{sec: Decay Timescale Selection}, we switch the order from the original [z, drop, x, drop, y, drop] to [x, drop, y, drop, z, drop]. The model is trained on all candidate features listed in Fig. \ref{fig:tau_efficiency}, and we scan over values of $\tau$ for each merger score definition, as shown in Fig. \ref{fig:tau_scan_projection}. The resulting $R^2$ curves remain stable, suggesting that the model performance is not sensitive to the choice of projection sequence, a reassuring result, as only one projection is actually observable for real clusters.

\subsection{Construction of richness-independent features}
\label{sec: Construction of Richness-Independent Features}
Since not all galaxy members can be reliably identified in observations, we construct a model that minimizes dependence on richness. Following the same procedures as in Section \ref{sec: Decay Timescale Selection}, we train the model using the following features: $\Delta v$, $\sigma_{r,1}$, $\sigma_{r,2}$, $\sigma_{v,1}$, $\sigma_{v,2}$, $\Delta\mathrm{bic}$, $\lambda_2/\lambda_1$, $z$, $f_m$, $\sigma_r$, and $\sigma_v$. These features are designed to be explicitly independent of the number of member subhalos.

\begin{figure*}[htbp]
    \centering
    \includegraphics[width=\textwidth]{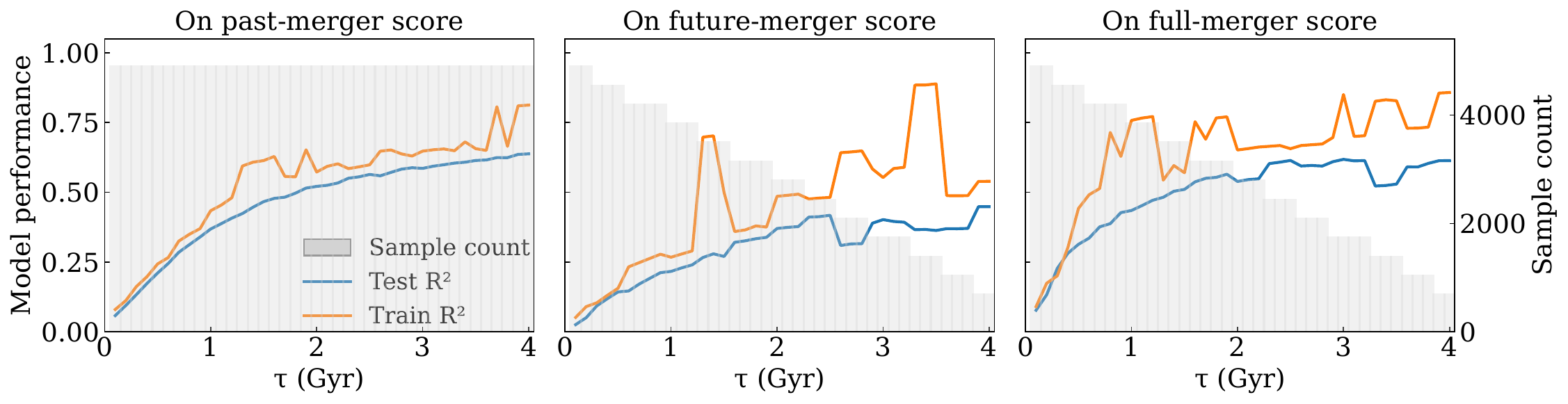}
    \caption{
    Model performance as a function of $\tau$, evaluated using three target-score definitions: past-merger (left), future-merger (middle), and full-merger (right).
    Orange and blue curves show the training and test $R^2$, respectively, and gray histograms indicate the sample count.
    The train--test split and sample selection follow Section~\ref{sec: Decay Timescale Selection} (grouping by halo ID and snapshot number) with rotation-based filtering.
    }
    \label{fig:tau_scan_discussion}
\end{figure*}

\begin{figure*}[htbp]
    \centering
    \includegraphics[width=\textwidth]{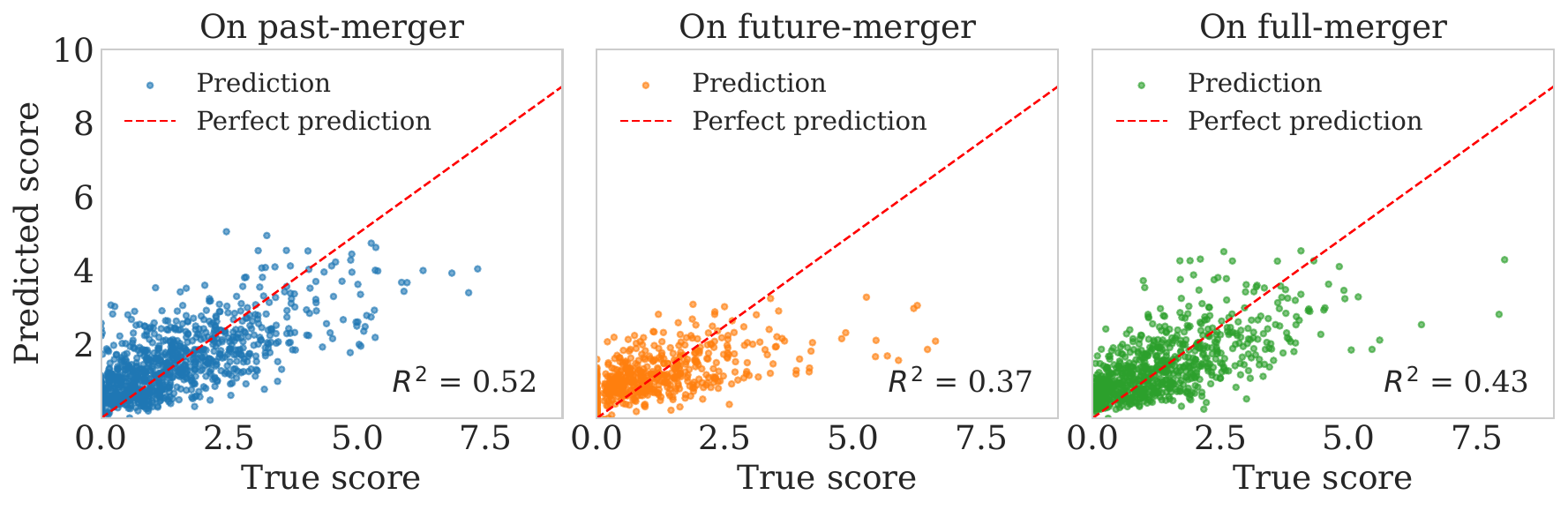}\\[2pt]
    \includegraphics[width=\textwidth]{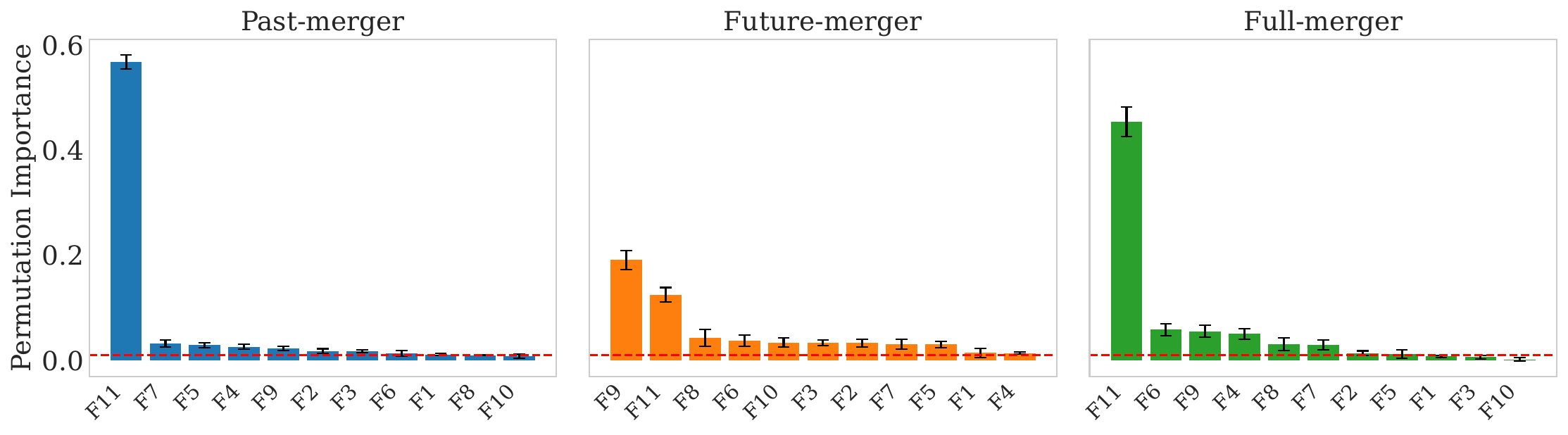}
    \caption{
    Richness-independent feature set results at selected $\tau$ values (Section~\ref{sec: Construction of Richness-Independent Features}): $\tau=2.0$~Gyr for the past- and future-merger scores and $\tau=1.0$~Gyr for the full-merger score.
    Top: predicted versus true merger scores for past-mergers (left), future-mergers (middle), and full-merger samples (right); the red dashed line indicates perfect prediction.
    Bottom: permutation feature importances for the corresponding models; bars show the mean importance and error bars show the standard deviation over 10 permutations. The red dashed line marks the 0.01 threshold used for feature selection. Features are labeled F1–F11, ranked by importance within each model. Their definitions are as follows: F1: $\Delta v$, F2: $\sigma_{r,1}$, F3: $\sigma_{r,2}$, F4: $\sigma_{v,1}$, F5: $\sigma_{v,2}$, F6: $\Delta\mathrm{bic}$, F7: $\lambda_2/\lambda_1$, F8: $z$, F9: $f_m$, F10: $\sigma_r$, F11: $\sigma_v$.
    }
    \label{fig:discussion_scatter_importance}
\end{figure*}

Fig. \ref{fig:tau_scan_discussion} to Fig. \ref{fig:discussion_scatter_importance} present the model results under this feature set. Specifically, we show the $\tau$-scan results for all three merger score definitions, scatter plots of predicted versus true scores at $\tau=2.0$ Gyr for past- and future-merger scores and $\tau=1.0$ Gyr for full-merger scores, as well as the corresponding feature importances. The test $R^2$ slightly decreases compared to the full-feature models (which include bic1 and bic2), but overall performance remains acceptable and interpretable.

\begin{figure}[ht!]
    \centering
    \includegraphics[width=\columnwidth]{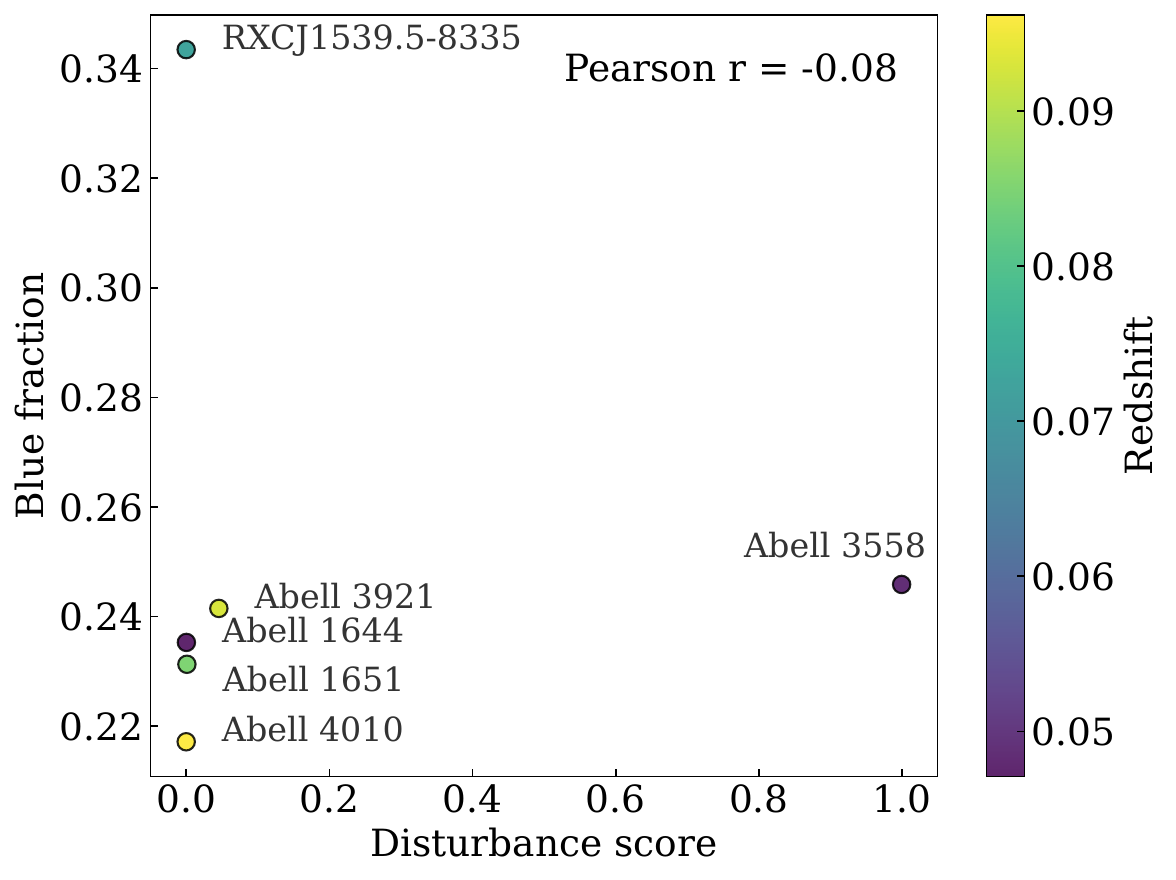}
    \caption{
        Correlation between the morphological disturbance score ($D_{\mathrm{comb}}$) and the fraction of blue galaxies in six LoVoCCS clusters.
        Each point represents one cluster, colored by redshift.
        A weak negative correlation is observed ($r = -0.08$).
        Cluster names are annotated; the redshift is encoded by color.
    }
    \label{fig:blue_frac_vs_dcomb}
\end{figure}
\begin{figure*}[ht!]
    \centering
    \includegraphics[width=\textwidth]{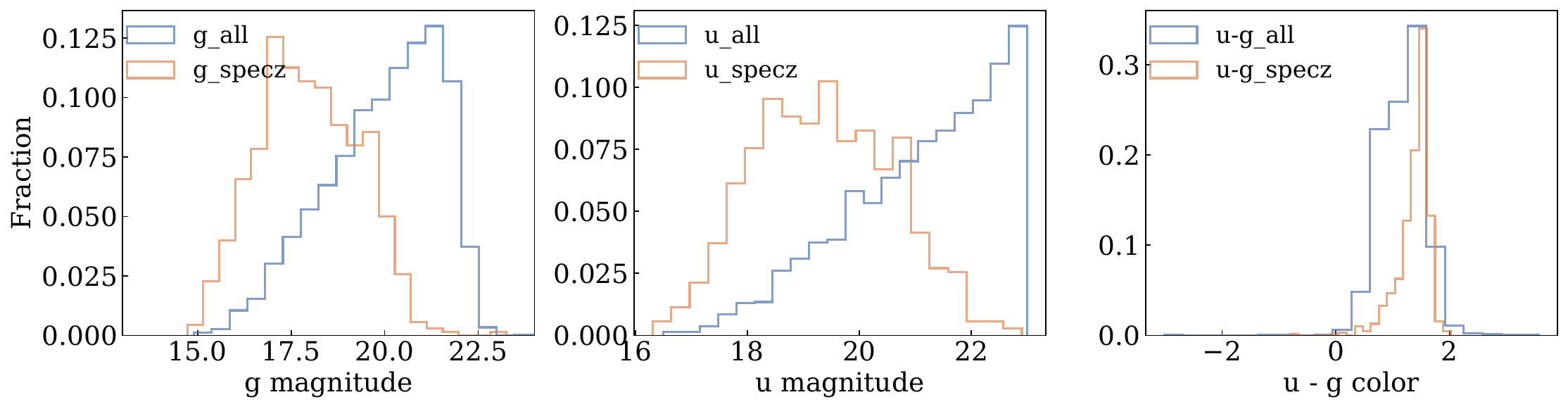}
    \caption{
        Comparison of photometric and spectroscopic galaxy member samples across the six LoVoCCS clusters used in our case study.
        Left: $g$-band magnitude distribution.
        Middle: $u$-band magnitude distribution.
        Right: $(u-g)$ color distribution.
        The histograms are normalized by sample size to show relative fractions.
        Galaxies with spectroscopic redshifts (orange) tend to be brighter and redder than the full photometric sample (blue), suggesting that spectroscopic samples preferentially include red sequence galaxies.
    }
    \label{fig:sample_distribution}
\end{figure*}
\begin{figure}[ht!]
    \centering
    \includegraphics[width=\columnwidth]{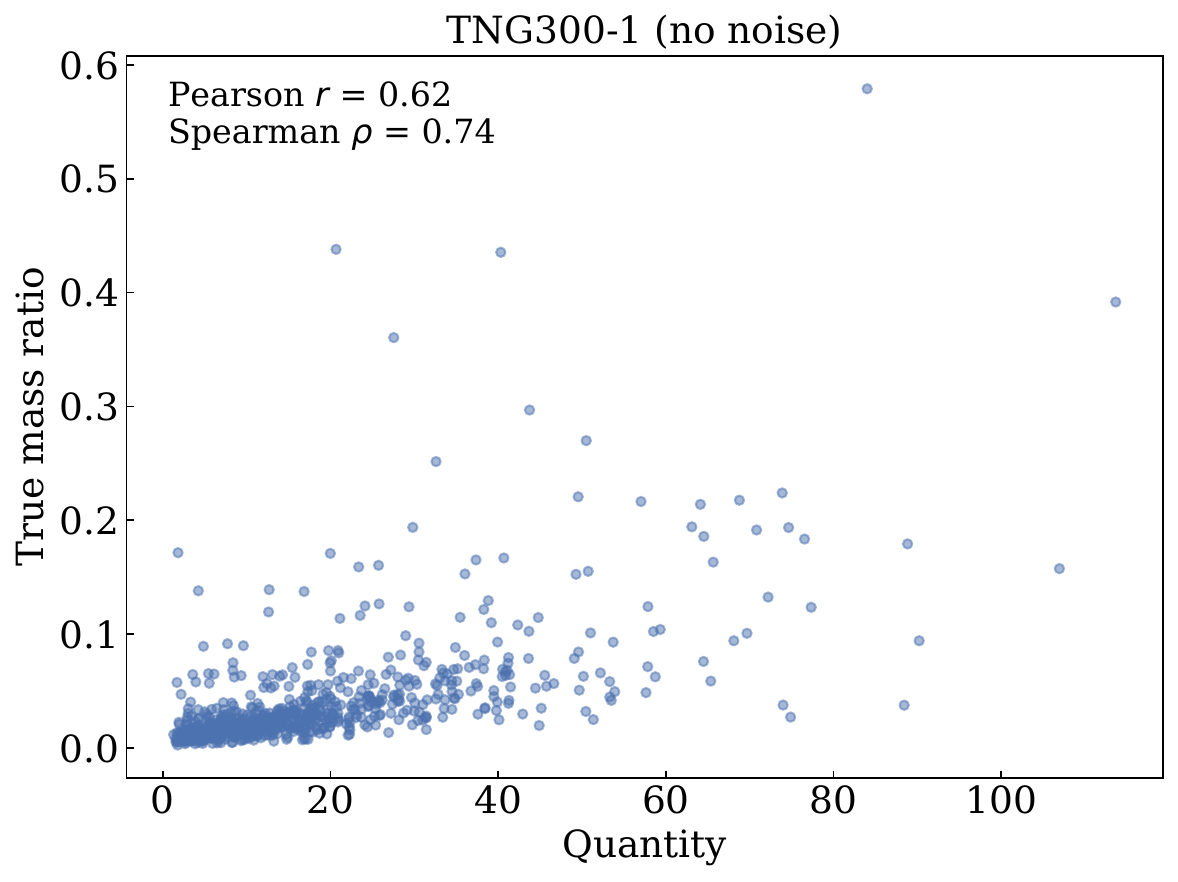}
    \caption{
        Correlation between the proxy quantity and the true mass ratio in the TNG300-1 simulation, under noise-free conditions. Each point corresponds to an FOF halo. The Pearson correlation coefficient is $r = 0.62$ and the Spearman rank correlation is $\rho = 0.74$, indicating a strong monotonic relationship.
    }
    \label{fig:mass_ratio_proxy_nonoise}
\end{figure}

\subsection{Case study: estimation of blue galaxy fraction}
\label{sec:blue_frac_case_study}

We match our LoVoCCS cluster catalog (with cluster information obtained from the NASA/IPAC Extragalactic Database) to the eROSITA X-ray survey \citep{erosita_full2024}, and obtain the morphological disturbance score $D_{\mathrm{comb}}$ from its associated morphology catalog \citep{2025_erosita_morphology}. Using a redshift-matching criterion of $\Delta z < 0.01$ and a maximum angular separation of 20 arcsec, we identify six matched clusters for this case study: Abell 4010, Abell 1651, Abell 1644, Abell 3558, Abell 3921, and RXCJ1539.5-8335. The matched sample is small, so we do not claim any statistically robust trend from this case study. We keep this section mainly as an example of how to apply our pipeline to observational datasets.

Due to the limited number of spectroscopic members, we rely on photometric redshifts for membership identification. We define members as galaxies within $R_{500c}$, provided by \citealt{erosita_full2024}, and within a photometric redshift interval $\Delta z = 0.05(1+z)$. A magnitude cut at $u, g < 23, 25$ is applied. We exclude sources with photometric odds $< 0.5$, extendedness $< 0.5$, or maximum photometric magnitude error $> 1.0$, where the odds are defined as a measure of the width of the photo-z probability distribution \citep{BPZ_paper}, reflecting the confidence in photometric redshift estimation. Red-sequence galaxies are identified by fitting a linear relation in the $(u-g)$ vs. $g$ color-magnitude space. Galaxies that lie 0.2 magnitudes below this fitted red sequence are classified as blue.

In Fig.~\ref{fig:blue_frac_vs_dcomb}, we present the relation between the disturbance score and the blue galaxy fraction for the six matched clusters. The correlation appears weak, which is likely due to the use of photo-z estimates in place of spectroscopic redshifts. More robust results could be obtained with an expanded dataset, for example, through additional X-ray observations of LoVoCCS clusters in the future. We also note that leakage of field galaxies from our photometric redshift technique will lessen the correlation.  Deeper and denser spectroscopic samples from the DESI (Dark Energy Spectroscopic Instrument) survey \citep{DESI_2024} and the upcoming Subaru Prime Focus Spectrograph (PFS) \citep{sabaru_PFS_2012} would improve the observational limits here.

For comparison, we also evaluate the galaxy sample distribution obtained using only spectroscopic members under the same selection criteria, with $\Delta z$ changed to $\Delta z = 0.01(1+z)$. We find that spectroscopic members tend to be systematically brighter and redder than the photometric sample, suggesting a potential selection bias when relying solely on spectroscopic redshifts that are currently available, as in Fig. \ref{fig:sample_distribution}.

To summarize, our method can in principle be applied to observational data, but several practical limitations need to be kept in mind. First, building an observational baseline requires a set of well-characterized 'standard' clusters with overlapping observational data (e.g., X-ray morphology measurements and sufficiently deep multi-band photometry), and such samples can be rare. Second, observational magnitude limits and incomplete membership identification mean that we do not recover all member galaxies, which can bias any member-based summary statistics, thus affecting GMM fits and the resulting BIC-derived features. Finally, using spectroscopic redshifts for membership selection can introduce selection bias, while using photometric redshifts is more complete but comes with larger uncertainties; both effects can dilute correlations and reduce model performance.

\subsection{Mass ratio estimator performance with future surveys}

Upcoming weak-lensing surveys such as Rubin/LSST \citep{guy_2025_15558559} and Roman/HLWAS \citep{akeson2019widefieldinfraredsurvey} are expected to achieve higher signal-to-noise ratios in aperture mass maps. This improvement will enhance the ability to identify multiple peaks, particularly in systems with lower merger mass ratios.

To assess the performance of our mass ratio estimator under such idealized conditions, we follow the procedure outlined in Section~\ref{sec: Mass Ratio Estimation}, but generate mock aperture mass maps without shape noise (i.e., background galaxies are assumed to have no intrinsic ellipticity). 

We include all halos with $M_{200c} > 5\times10^{13}M_{\odot}$, as well as halos with $1\times10^{13} < M_{200c} < 5\times10^{13}\,M_{\odot}$ if their mass ratio exceeds 0.1. We then examine the correlation between our derived proxy quantity (see Eq.~\ref{eq: rough_quantity}) and the true merger mass ratio in this noise-free scenario, as illustrated in Fig.~\ref{fig:mass_ratio_proxy_nonoise}.

Future observational advances should improve the reliability of merger identification. Upcoming redshift surveys and deeper/wider X-ray surveys will improve the detection of substructure and morphological asymmetries associated with mergers. Additionally, radio observations from MeerKAT \citep{2016_MeerKAT} and the upcoming SKA \citep{2009_SKA}, as well as mock radio maps from simulations, may help distinguish between past and future mergers. For instance, symmetric radio relics may emerge after the first pericenter passage and thus serve as potential indicators of post-merger systems.

\section{Conclusion}
We present a phase-space machine learning method for identifying major mergers in galaxy clusters, using features derived from the projected radius–velocity distribution of member galaxies in the TNG-Cluster simulation. Trained on a disturbance score built from merger history, the model effectively captures structural disturbances but lacks sensitivity to merger timing, failing to distinguish between past and future events.

To address this, we explored time-sensitive observables such as the curvature of the star formation rate (SFR) and the blue galaxy fraction. We find that the blue galaxy fraction decreases with increasing past-merger score, but shows no clear trend with future-merger score, suggesting its sensitivity to merger stage.

As a cross-check, we generated mock Chandra X-ray images and measured the BCG offset, which correlates strongly with the past-merger score but not with the future-merger score, supporting both the physical robustness of our score and the effectiveness of BCG offset as an indicator of past mergers.

Our results demonstrate that it is possible to determine the merger status of galaxy clusters based on the phase-space features and blue galaxy fraction.  As a practical matter, application of these methods will only improve as additional spectroscopic, X-ray, and weak lensing data becomes available in the next decade.

\begin{acknowledgements}
We thank the developers of the publicly available scientific software used in this work, including NumPy \citep{numpy}, Pandas \citep{reback2020pandas}, SciPy \citep{2020SciPy-NMeth}, Astropy \citep{The_Astropy_Collaboration_2022}, Matplotlib \citep{matplotlib}, scikit-learn \citep{pedregosa2018scikitlearnmachinelearningpython}, scikit-image \citep{scikit_image_van_der_Walt_2014}, pyXSIM \citep{pyXSIM}, SOXS \citep{SOXS}, and yt \citep{yt}.

We also acknowledge the IllustrisTNG project for providing the simulation data. The IllustrisTNG simulations were undertaken with compute time awarded by the Gauss 
Centre for Supercomputing (GCS) under GCS Large-Scale Projects GCS-ILLU and GCS-DWAR 
on the GCS share of the supercomputer Hazel Hen at the High Performance Computing 
Center Stuttgart (HLRS), as well as on the machines of the Max Planck Computing and 
Data Facility (MPCDF) in Garching, Germany.

This project used data obtained with the Dark Energy Camera (DECam), which was constructed by the Dark
Energy Survey (DES) collaboration. This work is based on observations at Cerro Tololo Inter-American Observatory,
NSF’s NOIRLab (NOIRLab Prop. ID 2019A-0308; PI: I. Dell’Antonio). 

This research is based on data obtained from the Astro Data Archive at NSF’s NOIRLab.

This research has made use of the NASA/IPAC Extragalactic Database (NED),
which is operated by the Jet Propulsion Laboratory, California Institute of Technology,
under contract with the National Aeronautics and Space Administration.

This work is based on data from eROSITA, the soft X-ray instrument aboard SRG, a joint Russian-German science mission supported by the Russian Space Agency (Roskosmos), in the interests of the Russian Academy of Sciences represented by its Space Research Institute (IKI), and the Deutsches Zentrum für Luft- und Raumfahrt (DLR). The SRG spacecraft was built by Lavochkin Association (NPOL) and its subcontractors, and is operated by NPOL with support from the Max Planck Institute for Extraterrestrial Physics (MPE). The development and construction of the eROSITA X-ray instrument was led by MPE, with contributions from the Dr. Karl Remeis Observatory Bamberg \& ECAP (FAU Erlangen-Nuernberg), the University of Hamburg Observatory, the Leibniz Institute for Astrophysics Potsdam (AIP), and the Institute for Astronomy and Astrophysics of the University of Tübingen, with the support of DLR and the Max Planck Society. The Argelander Institute for Astronomy of the University of Bonn and the Ludwig Maximilians Universität Munich also participated in the science preparation for eROSITA.

This research was conducted using computational resources and services at the Center for Computation and Visualization, Brown University.

The analysis code used in this work is publicly available at 
\url{https://github.com/KK-cloudhub/merger_trace}.
\end{acknowledgements}

\bibliographystyle{aa}      
\bibliography{ref} 
\end{document}